\documentclass[onecolumn,preprint,preprintnumbers,amsmath,amssymb,prb,apsrev4-2]{revtex4-2}
\usepackage{graphicx}
\usepackage{dcolumn}
\usepackage{bm}
\usepackage{SIunits}
\usepackage{verbatim}
\usepackage{placeins}
\usepackage{natbib}
\usepackage{amsmath}
\usepackage{amssymb}
\usepackage{longtable}
\usepackage{tabularx}
\usepackage{keyval}
\usepackage{multirow}
\usepackage[colorlinks,linkcolor=blue]{hyperref}

\begin{document}
\title{Universal quasi-degenerate orbital origin of two-dome phases in iron pnictide superconductors}
\author{Da-Yong Liu$^{1,2}$}
\email[]{dyliu@ntu.edu.cn}
\author{Zhe Sun$^{3}$}
\author{Feng Lu$^{4}$}
\author{Wei-Hua Wang$^{4}$}
\author{Liang-Jian Zou$^{2}$}
\email[]{zou@theory.issp.ac.cn}
\affiliation{1 School of physical science and technology, Nantong University, Nantong 226019, China}
\href{http://orcid.org/0000-0003-4370-473X}{ORCiD: 0000-0003-4370-473X}
\affiliation{2 Key Laboratory of Materials Physics, Institute of Solid State Physics, HFIPS, Chinese Academy of Sciences, Hefei 230031, China}
\affiliation{3 National Synchrotron Radiation Laboratory, University of Science and Technology of China, Hefei 230029, China}
\affiliation{4 Department of Electronic Science and Engineering, and Tianjin Key Laboratory of Photo-Electronic Thin Film Device and Technology, Nankai University, Tianjin 300350, China}

\date{\today}

\begin{abstract}
A series of experiments revealed that novel bipartite magnetic and superconducting (SC) phases widely exist in the phase diagrams of iron pnictides and chalcogenides. Nevertheless, the origin of the two-dome magnetic and SC phases in iron-based compounds remains unclear. Here we theoretically investigated the electronic structures, magnetic and SC properties of three representative iron-based systems, {\it i.e.} LaFeAsO$_{1-x}$H$_{x}$, LaFeAs$_{1-x}$P$_{x}$O and KFe$_{2}$As$_{2}$. We propose a unified quasi-degenerate orbital mechanism for the emergence of the two-dome parent magnetic/structural phase and the subsequent two-dome SC phase. It is found that the degenerate in-plane anisotropic $d_{xz/yz}$ orbitals dominate the first magnetic/structural and SC phases, while in-plane isotropic orbitals $d_{xy}$ or $d_{3z^{2}-r^{2}}$ with quasi-degeneracy originating from quasi-symmetry drive the emergence of the second magnetic/SC dome phase. Moreover, a matching rule of spin and orbital modes for SC pairing state is proposed in multi-orbital iron-based systems. These results imply an orbital-driven mechanism as well as an orbital-selective scenario, and shed light on the understanding of the multi-dome magnetic and SC phases in multi-orbital systems.
\end{abstract} 

\maketitle

\section{Introduction}
Iron-based superconductors display rich phase diagrams with antiferromagnetic (AFM) order \cite{Nature453-899,EPL83-27006}, orbital order (OO) \cite{PRL103-267001,PRB84-064435}, nematicity \cite{science329-824,nphys11-959,PRB92-085109} and superconductivity \cite{JACS130-3296,Nature453-761,PRL101-206404}. These competing or coexisting phases arise from the interplay among charge, spin, orbital and lattice degrees of freedom of iron-based compounds.
It is generally known that the AFM state is in close proximity to the superconducting (SC) state in the phase diagram of iron-based systems, indicating a spin-fluctuation mediated SC pairing state.
Moreover, it had already been realized that the multi-orbital features of electronic structures and OO play crucial roles in many different families of iron pnictides and chalcogenides \cite{nmat10-932,RMP85-849,PRL100-237003,PRL102-247001,NJP11-025021,PRL102-126401,PRL104-216405,PRL110-146402,RPB92-184512}, and the orbital selective Mott and SC physics were also suggested in some iron-based compounds \cite{PRL102-126401,PRL110-146402}. However, our understanding of the interplay between spin and orbital degrees of freedom is still accumulating, and more knowledge on its effects in SC pairing interactions, pairing symmetry, and phase diagrams in iron-based superconductors needs to be explored.

In the general phase diagram of iron-based superconductors, when the AFM state (here denoted as AFM-I) is suppressed by doping, the SC state (SC-I) occurs and its critical transition temperature (T$_{c}$) then reaches a maximum at the optimum doping but decreases with further doping, finally SC-I state is replaced by a normal state upon heavy doping, forming an ordinary single-dome feature. However, recent experiments have unambiguously revealed the presence of a second AFM (AFM-II) phase and an associated SC (SC-II) phase in the heavily doped region of iron pnictides and chalcogenides, forming a novel two-dome characteristic phase diagram. For instance, the parent compound LaFeAsO possesses a stripe AFM (SAFM) ground state associated with a weak $d_{xz}$ OO in the low-temperature phase and exhibits an orthorhombic to tetragonal structural phase transition upon doping or increasing temperature. Once substituting O with F, such a SAFM (AFM-I) is gradually suppressed and a SC-I dome emerges in LaFeAsO$_{1-x}$F$_{x}$ \cite{JACS130-3296}.
However, in addition to the first SC-I dome, with further increasing carrier concentration by means of hydrogen doping, recent experiments have shown that the second SC-II dome accompanied by bipartite magnetic phase emerges in LaFeAsO$_{1-x}$H$_{x}$, as depicted in the doping-dependent phase diagram (Fig.~\ref{Fig6}) \cite{ncomms3-943,nphys10-300}. Neutron diffraction experiments showed that the AFM-II phase lies around $x$=0.5, which also has SAFM with a large magnetic moment of 1.21 $\mu_{B}$ of Fe spin compared to 0.63 $\mu_{B}$ in LaFeAsO. Moreover, AFM-II possesses an unusual orthorhombic lattice distortion accompanied by the instability of As atoms that slightly move away from the center of the Fe square lattice in the low-temperature phase. Interestingly, it was very recently found that there also exist two SC domes in LaFeAsO$_{1-x}$F$_{x}$, and nuclear magnetic resonance (NMR) experiments showed a second orthorhombic phase for $x$$>$0.5 \cite{CPL32-107401}.

Accumulating evidence suggests that such new magnetic and SC phases take place in many iron-based SC families. For instance, a different doping-dependent phase diagram has been identified in LaFeAs$_{1-x}$P$_{x}$O \cite{JPSJ83-023707,JPSJ83-083702,PRB90-064504}. In doped K$_{1-x}$Fe$_{2-y}$Se$_{2}$, KFe$_{2}$As$_{2}$ and intercalated FeSe, the SC-II phase re-emerges under high pressure \cite{nature483-67,PRB91-060508,Sci.Rep.5-9477}. In addition, the SC-II dome has also been observed in the K-doped FeSe thin films grown on SiC substrate \cite{PRL116-157001}. On the other hand, high pressure can induce a new AFM phase in pure FeSe \cite{PRB93-094505}. Instead, some systems only have single-dome AF/SC phase character, thus the two-dome phase is attributed to be an accidental phenomenon. How do the unusual phenomena emerge? Although more evidence of the existence of the AFM-II and SC-II phases has been found in various iron-based compounds, their origin and nature still remain unclear.

In this work, we have combined first-principle electronic structure calculation and analytical methods including mean-field method and multi-orbital random phase approximation (RPA) to investigate the electronic, magnetic and SC properties of three representative systems with novel two-dome phases, {\it i.e.} LaFeAsO$_{1-x}$H$_{x}$, LaFeAs$_{1-x}$P$_{x}$O and KFe$_{2}$As$_{2}$. We found that the second parent AFM-II or structural phase originates from the instability of the quasi-orbital degeneracy. Moreover, we also demonstrate that the in-plane anisotropic/isotropic orbitals dominate the SC-I/SC-II phase, respectively, implying an orbital-selective scenario.
Our results have shown that the novel AFM-II and SC-II phases indeed have different orbital physics in contrast to the widely studied conventional AFM-I and SC-I phases.
The rest of this paper is organized as follows: the model Hamiltonian and method are described in Sec. II; results and discussion, including three typical two-dome SC systems, LaFeAsO$_{1-x}$H$_{x}$, LaFeAs$_{1-x}$P$_{x}$O and KFe$_{2}$As$_{2}$, are presented in Secs. III, and a unified orbital-driven scenario for the two-dome phase and an orbital-spin mode matching rule for pairing state are also depicted; last are the remarks and conclusions.

\section{Model and Method}
To model the evolution of the two-dome phase diagram on the doping amount or pressure, we first perform electronic structure calculations, then extract the tight-binding models, and finally theoretically analyze the magnetic and SC properties based on the multi-orbital Hubbard model including Coulomb interaction.
The electronic structures calculations are performed by the full potential linearized augmented-plane-wave (FP-LAPW) scheme based on density functional theory (DFT) implemented with WIEN2K \cite{WIEN2K}. The generalized gradient approximation (GGA) by Perdew, Burk, and Ernzerhof (PBE) \cite{PBE} is used to take into account the exchange and correlation effects.

Considering that the parent phases LaFeO$_{1-x}$H$_{x}$ ($x$=0.5) and LaFeAs$_{1-x}$P$_{x}$O ($x$=0.5) are both the AFM ordered states according to the experimental data \cite{nphys10-300,JPSJ83-023707}, the supercell structures are adopted to simulate the H and P doping effects, which is a good starting point for $x$=0.5 case. Moreover, since the crystal field dependent on local tetrahedral ligand environment is seriously concerned, the supercell approach is a good treatment of doping in comparison with the virtual crystal approximation (VCA). 
Consequently, the supercell structures can be used for simulating the doped LaFeAsO$_{1-x}$H$_{x}$ and LaFeAs$_{1-x}$P$_{x}$O systems at $x$=0.5, as displayed in Fig. S2 and Fig. S5 of the Supplementary Material \cite{SM}.

The maximally localized Wannier functions (MLWFs) implemented with WANNIER90 \cite{CPC178-685} and WIEN2WANNIER \cite{CPC181-1888} are used to construct the tight-binding model $H_{0}$:
\begin{eqnarray}
H_{0}=\sum_{\substack{i,j\\ \alpha,\beta,\sigma}}t_{ij}^{\alpha\beta}c_{i\alpha \sigma }^{\dagger }{{c}_{j\beta \sigma }}-\mu\sum_{\substack{i\alpha\sigma}}n_{i\alpha\sigma}.
\label{eqtb}
\end{eqnarray}
The Wannier functions fitting band structures of three representative LaFeAsO$_{1-x}$H$_{x}$, LaFeAs$_{1-x}$P$_{x}$O and pressured-KFe$_{2}$As$_{2}$ systems are presented in the Supplementary Material \cite{SM}.
In order to analyse the orbital features of Fe-3$d$, crystal field parameters, {\it i.e.} on-site energies $\varepsilon_{\alpha}$ ($t_{ij}^{\alpha\beta}$ with $i$=$j$ and $\alpha$=$\beta$), are extracted through the projected Wannier functions of 3$d$ orbitals onto the atomic orbitals.

On the other hand, the electronic correlation effect is treated within the mean-field and multi-orbital RPA approaches. The electronic interaction part of the multi-orbital Hamiltonian $H_{I}$ term reads,
\begin{eqnarray}
  H_{I} &=& U\sum_{\substack{i, \alpha}}n_{i\alpha\uparrow}n_{i\alpha\downarrow}
  +U^{'}\sum_{\substack{i\\ \alpha\ne\beta}}n_{i\alpha\uparrow}n_{i\beta\downarrow}
  +(U^{'}-J_{H})
   \nonumber\\
  &&\times\sum_{\substack{i,\sigma\\ \alpha<\beta}}n_{i\alpha\sigma}n_{i\beta\sigma}
  -J_{H}\sum_{\substack{i\\ \alpha\ne\beta}}
  C_{i\alpha\uparrow}^{\dag}C_{i\alpha\downarrow}C_{i\beta\downarrow}^{\dag}C_{i\beta\uparrow}
  \nonumber\\
  &&+J_{H}\sum_{\substack{i\\ \alpha\ne\beta}}
  C_{i\alpha\uparrow}^{\dag}C_{i\alpha\downarrow}^{\dag}C_{i\beta\downarrow}C_{i\beta\uparrow},
\end{eqnarray}
where $U$($U^{'}$) denotes the intra-(inter-)orbital Coulomb repulsion interaction and $J_{H}$ the Hund's rule coupling. Considering the symmetry of the system, we adopt $U^{'}$=$U-2J_{H}$.

To explore the interplays among the charge, spin and orbital degrees of freedom including the electronic correlation in AFM-I and AFM-II parent phases, we consider the five-orbital Hubbard models $H=H_{0}+H_{I}$ to investigate the electronic and magnetic properties for different doping iron-based systems. Within the mean-filed approximation, we define the orbital-resolved occupation and magnetic order parameters $n_{\alpha}$ and $m_{\alpha}$ with $\alpha$=$d_{xz}$, $d_{yz}$, $d_{xy}$, $d_{x^{2}-y^{2}}$ and $d_{3z^{2}-r^{2}}$ as
\begin{eqnarray}
 n_{\alpha}=\sum_{\substack{k,\sigma}}\langle c_{k,\alpha,\sigma}^{\dag}c_{k,\alpha,\sigma} \rangle, m_{\alpha}=\sum_{\substack{k,\sigma}}\sigma\langle c_{k+\mathbf{Q},\alpha,\sigma}^{\dag}c_{k,\alpha,\sigma} \rangle
 \label{eq.3}
\end{eqnarray}
where $\mathbf{Q}$ is the magnetic vector and $\sigma$=$\pm$1. Then we can obtain the magnetic moment $m=\sum_{\alpha}m_{\alpha}$ and total particle number $n=\sum_{\alpha}n_{\alpha}$ through the self-consistently calculations.

Moreover, we study the effect of spin and orbital fluctuations on SC pairing using the multi-orbital RPA method.
The bare susceptibility is given by the formula \cite{NJP11-025016,RPP74-124508,RMP84-1383}
\begin{eqnarray}
  \chi_{0}^{l_{1}l_{2}l_{3}l_{4}}(\mathbf{q},\omega)&=&-\frac{1}{N}\sum_{\substack{\mathbf{k},\mu\nu}}
  [f(\varepsilon_{\mu}(\mathbf{k}+\mathbf{q}))-f(\varepsilon_{\nu}(\mathbf{k}))] \\
  &&\times\frac{a_{\nu}^{l_{4}}(\mathbf{k})a_{\nu}^{l_{2},*}(\mathbf{k})a_{\mu}^{l_{1}}(\mathbf{k+q})a_{\mu}^{l_{3},*}(\mathbf{k+q})}
  {\omega+\varepsilon_{\mu}(\mathbf{k}+\mathbf{q})-\varepsilon_{\nu}(\mathbf{k})+i\eta} \nonumber
  \label{eq-chi0}
\end{eqnarray}
The RPA susceptibility is expressed as
\begin{eqnarray}
  \chi_{c(s)}^{RPA}(\mathbf{q}, \omega) &=& \chi_{0}(\mathbf{q},\omega)[\mathbb{I}\pm\Gamma_{c(s)}\chi_{0}(\mathbf{q},\omega)]^{-1},
\end{eqnarray}
where $\chi_{0}$ defined in Eq.~\ref{eq-chi0} is the bare susceptibility, and the nonzero components of the interaction matrices $\Gamma_{c}$ and $\Gamma_{s}$ are given as $(\Gamma_{c})_{aa,aa}=U$, $(\Gamma_{c})_{aa,bb}=2U'-J_{H}$, $(\Gamma_{c})_{ab,ab}=-U'+2J_{H}$, $(\Gamma_{c})_{ab,ba}=J_{H}$ and $(\Gamma_{s})_{aa,aa}=U$, $(\Gamma_{s})_{aa,bb}=J_{H}$, $(\Gamma_{s})_{ab,ab}=U'$, $(\Gamma_{s})_{ab,ba}=J_{H}$ with orbitals $a\neq b$.
The singlet orbital vertex function $\Gamma_{l_{1}l_{2}l_{3}l_{4}}$ is given by
\begin{eqnarray}
  \Gamma_{l_{1}l_{2}l_{3}l_{4}}(\mathbf{k},\mathbf{k'},\omega)=[\frac{3}{2}U^{s}\chi_{s}^{RPA}(\mathbf{k-k'}, \omega)U^{s}+\frac{1}{2}U^{s} \nonumber\\
  -\frac{1}{2}U^{c}\chi_{c}^{RPA}(\mathbf{k-k'},\omega)U^{c}+\frac{1}{2}U^{c}]_{l_{1}l_{2}l_{3}l_{4}}. \nonumber
\end{eqnarray}
\begin{eqnarray}
  \Gamma_{ij}(\mathbf{k},\mathbf{k'})&=& Re[\sum_{\substack{l_{1}l_{2}l_{3}l_{4}}}a_{\nu_{i}}^{l_{2},*}(\mathbf{k})a_{\nu_{i}}^{l_{3},*}(\mathbf{-k}) \nonumber\\
  &&\Gamma_{l_{1}l_{2}l_{3}l_{4}}(\mathbf{k},\mathbf{k'},\omega=0)
  \times a_{\nu_{j}}^{l_{1}}(\mathbf{k'})a_{\nu_{j}}^{l_{4}}(\mathbf{-k'})]. \nonumber
\end{eqnarray}
\begin{eqnarray}
 &&-\frac{\sum_{\substack{ij}}\oint_{C_{i}}\frac{d\mathbf{k}_{\|}}{v_{F}(\mathbf{k})}\oint_{C_{j}}\frac{d\mathbf{k}_{\|}^{'}}{{v_{F}(\mathbf{k'})}}\Delta(\mathbf{k})\Gamma_{ij}(\mathbf{k},\mathbf{k'})\Delta(\mathbf{k'})}{(2\pi)^{2}\sum_{\substack{i}}\oint_{C_{i}}\frac{d\mathbf{k}_{\|}}{v_{F}(\mathbf{k})}[\Delta(\mathbf{k})]^{2}}\nonumber\\
 &&=\lambda[\Delta(\mathbf{k})]
\end{eqnarray}
where the Fermi velocity $v_{F}(\mathbf{k})=|\nabla_{\mathbf{k}}E_{\nu}(\mathbf{k})|$. With these equations, one could study the SC pairing strengths of the multi-orbital systems.

\section{Results and discussion}
Here we aim to clarify the origin of the two-dome AFM and SC phases in LaFeAsO$_{1-x}$H$_{x}$, LaFeAs$_{1-x}$P$_{x}$O, and KFe$_{2}$As$_{2}$ compounds. We first study the electronic, magnetic and SC properties of LaFeAsO$_{1-x}$H$_{x}$ using the {\it first-principles} method, projected Wannier functions, random phase approximation (RPA) and mean-field approximation based on five-orbital Hubbard models. We found that the quasi-degeneracy of electronic states with different orbital symmetries plays the key role in the emergence of the AFM-II phase as well as structural phase transition, and that the active isotropic orbital governs the SC-II phase, leading to two SC domes in the phase diagrams of these iron pnictides and chalcogenides. In LaFeAsO$_{1-x}$H$_{x}$, the quasi-degenerate $d_{xy}$/$d_{xz/yz}$ orbitals drive the emergence of the AFM-II phase, and the in-plane isotropic $xy$-orbital dominates an orbital-selective SC pairing state. In LaFeAs$_{1-x}$P$_{x}$O, one of the quasi-degenerate orbitals, the in-plane isotropic $d_{3z^{2}-r^{2}}$ orbital, is responsible for the AFM-II phase. In KFe$_{2}$As$_{2}$, a quasi-degenerate $d_{xy}$/$d_{xz/yz}$ orbital instability leads to a structural phase transition from a tetragonal phase into a collapsed tetragonal (CT) phase under pressure.

\subsection{LaFeAsO$_{1-x}$H$_{x}$}
\subsubsection{Tight-binding model and electronic structures}
In order to compare our theoretical electronic structure results with the experimental data, we use the experimental parameters \cite{nphys10-300} of the crystal structure of LaFeAsO$_{1-x}$H$_{x}$ ($x$=0.5), as seen in Fig. S2 (Supplementary Material \cite{SM}). The space groups of the low-temperature and high-temperature phases are $C_2$ (No. 5) and $P$-$4m2$ (No. 115), respectively.

We first adopt the high-temperature structure to calculate its electronic structures, then employ the Wannier function technique to obtain the projected Wannier functions of Fe-3$d$ orbitals onto the atomic orbitals for LaFeAsO$_{1-x}$H$_{x}$ at $x$=0 and 0.5 (See Figs. S1 and S3 of the Supplementary Material for details \cite{SM}), respectively. The on-site energies of Fe-3$d$ orbitals are listed in Table.~\ref{Tab1}.
%
\begin{table}[htbp]
\caption{On-site energies of Fe-3$d$ orbitals for LaFeAsO$_{1-x}$H$_{x}$ measured in units of eV.}
\label{Tab1}
\begin{tabular}{lccc}
\hline\hline
\multirow{1}{6cm}{LaFeAsO$_{1-x}$H$_{x}$}
$xz$/$yz$ & $xy$ & $x^{2}-y^{2}$ & $3z^{2}-r^{2}$ \\
\hline
\multirow{1}{6cm}{$x$=0 (LaFeAsO) \cite{PRL101-087004}}
$0.0$&$0.16$&$-0.34$&$-0.21$  \\
\hline
\multirow{1}{6cm}{$x$=0.5}
$\mathbf{0.0}$&$\mathbf{0.04}$&$-0.24$&$-0.19$ \\
\hline\hline
\end{tabular}
\end{table}
In LaFeAsO$_{1-x}$H$_{x}$, the substitution of H$^{-}$ ions for O$^{2-}$ ions in the LaO$_{1-x}$H$_{x}$ layer will have two effects on the FeAs layer. One is the electron doping effect, the other is the chemical pressure (compression) effect of LaO$_{1-x}$H$_{x}$ layer on the FeAs layer due to the much larger H$^{-}$ ion radius. The former changes the electron occupation of Fe ions from $n$=6 ($x$=0) to $n$=6.5 ($x$=0.5), while the latter changes the crystal field of Fe-3$d$ orbitals in FeAs tetrahedra.
It is found that with increasing doping concentration ($x$), the on-site energy of $d_{xy}$ orbital, $\varepsilon_{xy}$, is closer to that of $d_{xz}$/$d_{yz}$ orbitals $\varepsilon_{xz/yz}$, as given in Table.~\ref{Tab1}.
Especially in $x$=0.5 case, the $d_{xz}$/$d_{yz}$ orbitals are nearly degenerate with the $d_{xy}$ orbital. Here we define this crystal field configuration as quasi-degeneracy, {\it i.e.} the on-site energies $\varepsilon_{\alpha}$ are equal but the bandwidths $W_{\alpha}$ are different, in comparison with the fully degenerate with the equal on-site energies and bandwidths, as displayed later in Fig.~\ref{Fig15}.
Notice that although the $d_{xy}$ and $d_{xz}$/$d_{yz}$ orbitals are not fully degenerate at $x$=0.5, considering that the evolutions of the doping, substitution or pressure are successive processes, one could find that there exists a critical doping concentration $x_{c}$ or a critical pressure $P_{c}$, where the three orbitals are fully degenerate near $x$=0.5.
In addition, the $d_{x^{2}-y^{2}}$ orbital is also close to the $d_{3z^{2}-r^{2}}$ orbital, resulting in approximate cubic symmetry with degenerated $t_{2g}$ ($d_{xz}$/$d_{yz}$/$d_{xy}$) and $e_{g}$ ($d_{x^{2}-y^{2}}$/$d_{3z^{2}-r^{2}}$) orbitals at $x$=0.5.
It means that the compression effect on FeAs tetrahedra causes the system to enter a higher quasi-symmetry in the vicinity of cubic one.

\begin{figure}[htbp]
\hspace*{-2mm}
\centering
\includegraphics[trim = 0mm 0mm 0mm 0mm, clip=true, width=0.65 \columnwidth]{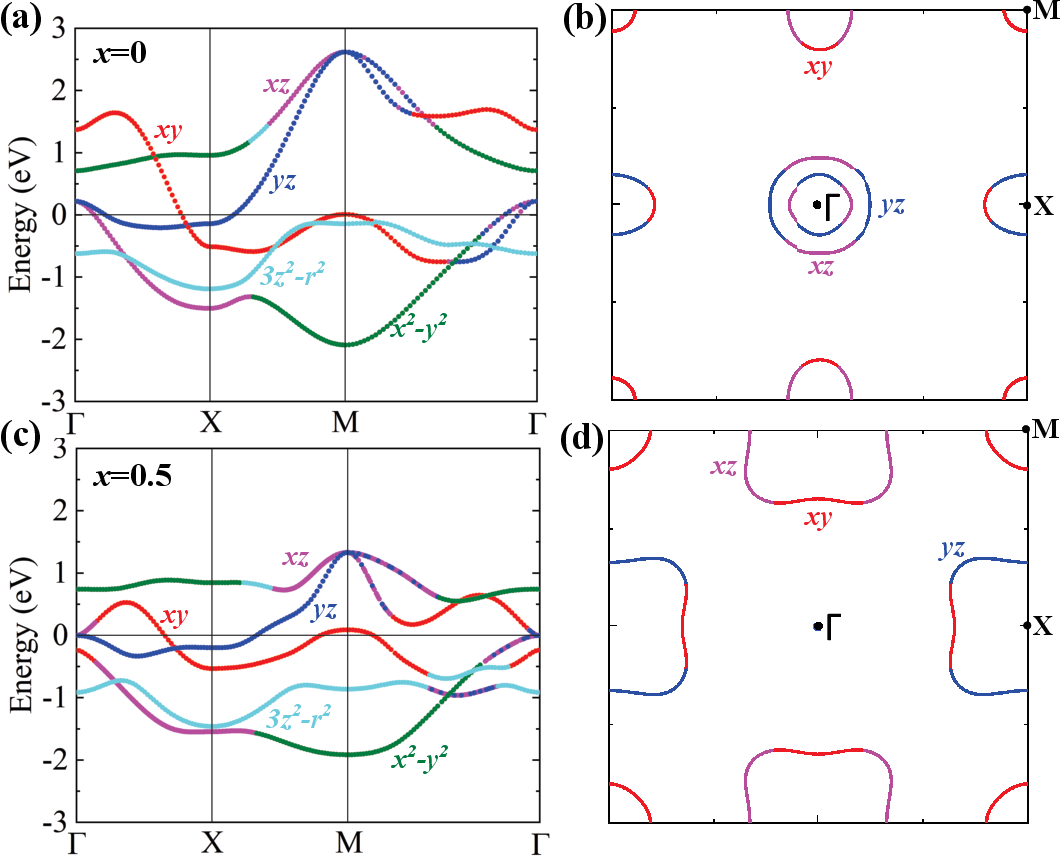}
\caption{(Color online) Band structures and Fermi surfaces for LaFeAsO ($x$=0) (a) and (b), and LaFeAsO$_{1-x}$H$_{x}$ ($x$=0.5) (c) and (d), respectively, with colors indicating majority orbital character.}
\label{Fig1}
\end{figure}

Based on the projected Wannier functions, we construct five-orbital tight-binding models $H_{0}$ of Fe-3$d$ bands by fitting the band structures of $x$=0 ($n$=6) and $x$=0.5 ($n$=6.5), respectively, as seen in Eq.~\ref{eqtb}. Our five-orbital tight-binding models including band structures and Fermi surfaces are consistent with that in Refs.~\cite{PRL113-027002,NJP12-073030}. The obtained band structures and Fermi surfaces are shown in Fig.~\ref{Fig1} (a), (c) and Fig.~\ref{Fig1} (b), (d), respectively.
Notice in particular that the three $d_{xz}$, $d_{yz}$, $d_{xy}$ orbitals are active ones because they are very close to the Fermi level $E_{F}$.
It is worth noting that in order to verify the reliability of doping calculations, our supercell calculation results are also compared with those of the VCA in LaFeAsO$_{1-x}$H$_{x}$ ($x$ $=$ 0.5) system \cite{RRB94-224511}, and found a consistent electronic band structure.

The five-orbital tight-binding models of Fe-3$d$ bands are constructed for $x$=0 ($n$=6) and $x$=0.5 ($n$=6.5), respectively. The orbital-resolved band structures are shown in Fig.~\ref{Fig1}(a) and~\ref{Fig1}(c), respectively. Here orbitals $d_{xz}$, $d_{yz}$, $d_{xy}$, $d_{x^{2}-y^{2}}$, $d_{3z^{2}-r^{2}}$ are denoted as 1, 2, 3, 4, 5, respectively. 
It is well known that the iron-based superconductor exhibits multi-orbital characteristic electronic structure. It is worth noting that, all five Fe-3$d$ orbitals indeed contribute to the Fermi surface with different weights. The band structure and Fermi surface are plotted using color to delineate the dominant orbital character, which possesses the greatest weight, as displayed in Fig.~\ref{Fig1}. 
In fact, both the dominant degenerate/quasi-degenerate active orbitals and individual subdominant active orbitals contribute significantly to the electronic state at the Fermi level $E_{F}$.

The interplay of spin-orbital orderings or fluctuations are analyzed in the multi-orbital Hubbard models by both the mean-field approximation and RPA.
To reveal the origin of the AFM(II)-SC(II) phase, the crystal field splittings of Fe-3$d$ orbitals, {\it i.e.} on-site energies, are analyzed. It is worth noting that, due to different orbital components and weights in the two magnetic phases AFM-I and AFM-II, a single rigid-band tight-binding model could not describe the entire $T$-$x$ phase diagram. In order to depict the low-energy physics in LaFeAsO$_{1-x}$H$_{x}$, we thus adopt two sets of tight-binding parameters for $x$=0 \cite{PRL101-087004} and 0.5 to describe the AFM(I)-SC(I) and AFM(II)-SC(II) phases, respectively. In LaFeAsO$_{1-x}$H$_{x}$ ($x$=0.5), $d_{xz}$/$d_{yz}$ are almost degenerate with $d_{xy}$, {\it i.e. quasi-degenerate}, with bandwidth $W_{xz/yz}>W_{xy}$. In the following, combined with the analysis of magnetic, orbital and superconducting pairing properties, we argue that the occurrence of the quasi-degeneracy of $d_{xy}$ and $d_{xz}$/$d_{yz}$ orbitals drives the emergence of the AFM-II and SC-II phases.

\subsubsection{Magnetic exchange interaction}
To understand the nature of the AFM-II parent phase, we calculate the total energy of various low-temperature magnetic configurations of LaFeAsO$_{1-x}$H$_{x}$ at $x$=0.5 by the {\it first-principle} DFT calculations. 
During the DFT calculation, we use the low-temperature orthorhombic structure with lattice parameter $a$ $>$ $b$ for the AFM-II phase (LaFeAsO$_{0.5}$H$_{0.5}$) according to the experimental data.
Several magnetic structures are considered, including nonmagnetic (NM), ferromagnetic (FM), N$\acute{e}$el AFM (NAFM) and striped AFM (SAFM) states, with SAF1 (SAF2) denoting the SAFM structure with the AFM coupling along the $a$ ($b$) axis and the FM coupling along the $b$ ($a$) axis for the orthorhombic structure with $a$ $>$ $b$, respectively. Based on the results of these electronic structures calculations, we find that the SAF1 state is the magnetic ground state, the same as that of LaFeAsO, in agreement with the neutron diffraction experiment~\cite{nphys10-300}.

We then construct an effective $J_{1a}$-$J_{1b}$-$J_{2}$ Heisenberg model to describe the magnetic interactions and low-energy physics of the AFM-II phase in LaFeAsO$_{1-x}$H$_{x}$ at $x$=0.5:
\begin{equation}
\label{eq.1}
\begin{aligned}
H=&J_{1a}\sum_{\substack{<ij>_{a}}}\vec{s}_{i}\cdot\vec{s}_{j}
+J_{1b}\sum_{\substack{<ij>_{b}}}\vec{s}_{i}\cdot\vec{s}_{j}\\
\nonumber
&+J_{2}\sum_{\substack{<<ij>>}}\vec{s}_{i}\cdot\vec{s}_{j},
\end{aligned}
\end{equation}
where $J_{1a}$, $J_{1b}$ and $J_{2}$ are the nearest-nearest-neighbor (N.N.) along the $a$- and $b$-axes, and next nearest-nearest-neighbor (N.N.N.) spin exchange parameters with spin $s$ of Fe ions, respectively, as shown in the inset of Fig.~\ref{Fig6}. The spin exchange parameters can be fitted by the differences of total energies per Fe ions between different magnetic structures, and are listed in what follows,
\begin{equation}
\label{eq.2}
\begin{aligned}
\Delta E(NAF-FM)&=&&-2(J_{1a}+J_{1b})s^{2} \\
\nonumber
\Delta E(NAF-SAF1)&=&&-2(J_{1b}-2J_{2})s^{2} \\
\nonumber
\Delta E(NAF-SAF2)&=&&-2(J_{1a}-2J_{2})s^{2}.
\end{aligned}
\end{equation}

According to the {\it first-principles} calculations, the stable magnetic ground state in the AFM-II phase is SAFM in LaFeAsO$_{0.5}$H$_{0.5}$, in agreement with the neutron diffraction experiment \cite{nphys10-300}. Based on the total energy differences among different magnetic configurations and the fitting by a $J_{1a}$-$J_{1b}$-$J_{2}$ Heisenberg model, we estimate the spin exchange parameters: the N.N. spin coupling $J_{1a}$=7.8 meV/$s^{2}$, $J_{1b}$=4.7 meV/$s^{2}$, and the N.N.N. coupling $J_{2}$=31.4 meV/$s^{2}$ in LaFeAsO$_{0.5}$H$_{0.5}$. $J_{2}$ is much larger than $J_{1a}$ and $J_{1b}$, {\it i.e.} $J_{1}$ $\ll$ $J_{2}$, which is very different from the case in the AFM-I ($J_{1}$ $>$ $J_{2}$). 
Within the DFT calculation, we observe OO with large $d_{xy}$ orbital occupation at Fermi level $E_{F}$, as displayed in Fig. S4 (Supplementary Material \cite{SM}). As is well known, considering the orbital degree of freedom, the magnetic coupling strength with $J$($\tau_{i}$, $\tau_{j}$) is orbital dependent in the Kugel-Khomski model with orbital operator $\tau$. As a consequence, in the first AFM (AFM-I) parent phase, $n_{xz}$ $>$ $n_{yz}$ with $d_{xz}$ OO leads to $J_{1a}$ $>$ $J_{1b}$, while in the second AFM parent (AFM-II) phase, $n_{xy}$ $>$ $n_{xz}$ $>$ $n_{yz}$ leads to $J_{2}$ $>$ $J_{1a}$ $>$ $J_{1b}$, implying existence of strong $d_{xy}$ and weak $d_{xz}$ OO.
This strong N.N.N. magnetic interaction $J_{2}$ or anomalous magnetic coupling constant ratio $J_{2}$/$J_{1}$ $>$ 1 indicates the possible existence of strong OO orientating the diagonal direction of Fe square lattice.
Meanwhile, the SAF1 state is more stable with the slightly different exchange coupling parameters $J_{1a}$ $>$ $J_{1b}$, indicating the existence of the weak $d_{xz}$ OO corresponding to the orthorhombic lattice parameter $a$ $>$ $b$.

\subsubsection{Orbital order: mean-field method}
\begin{figure}[htbp]
\hspace*{-2mm}
\centering
\includegraphics[trim = 0mm 0mm 0mm 0mm, clip=true, width=0.65 \columnwidth]{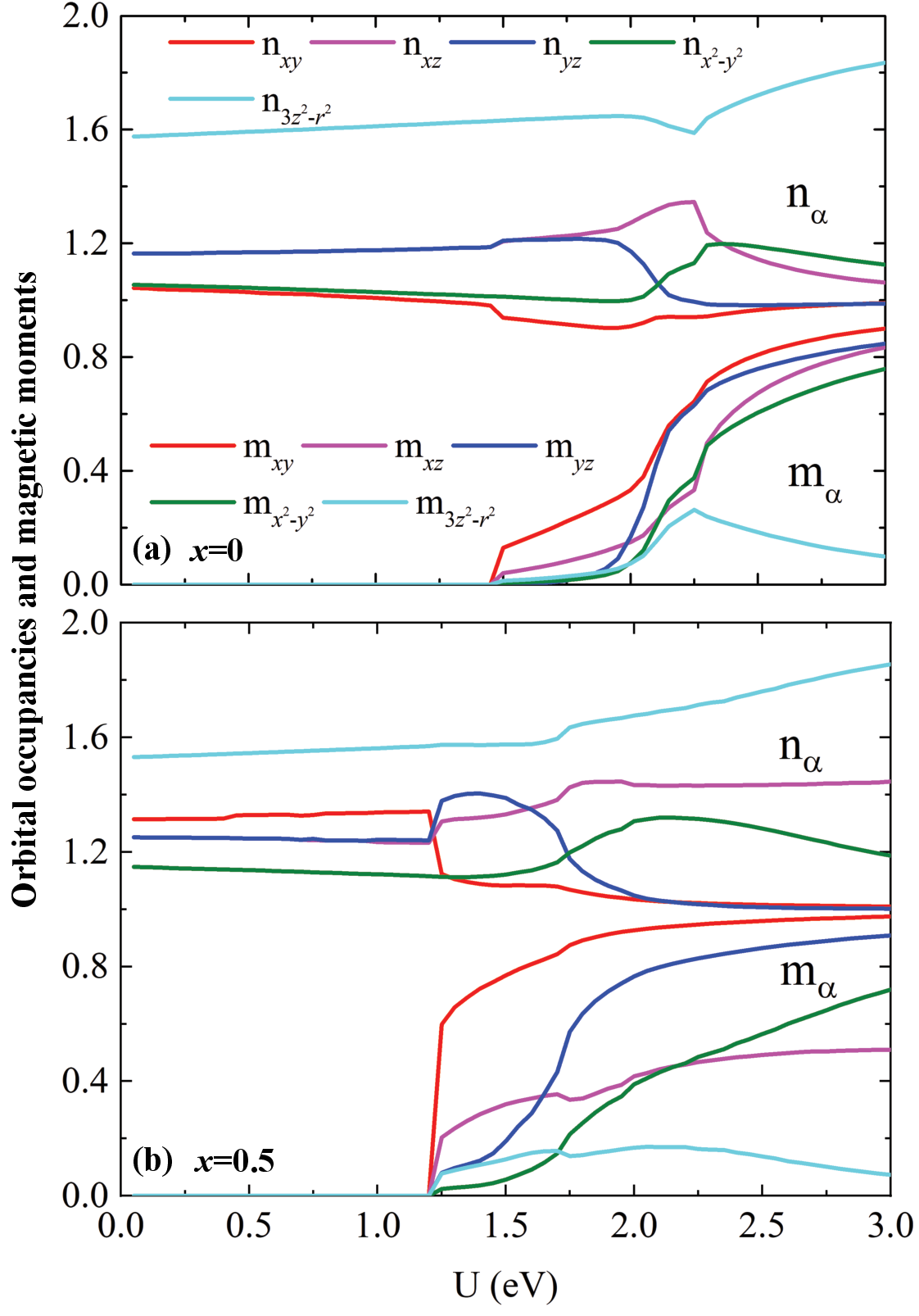}
\caption{(Color online) Orbital occupancies and magnetic moments dependence on Coulomb interaction $U$ within the five-orbital Hubbard model for (a) LaFeAsO ($x$=0) and (b) LaFeAsO$_{1-x}$H$_{x}$ ($x$=0.5) with parameter $J_{H}$=0.1$U$.}
\label{Fig2}
\end{figure}

\begin{figure}[htbp]
\hspace*{-2mm}
\centering
\includegraphics[trim = 0mm 0mm 0mm 0mm, clip=true, width=0.65 \columnwidth]{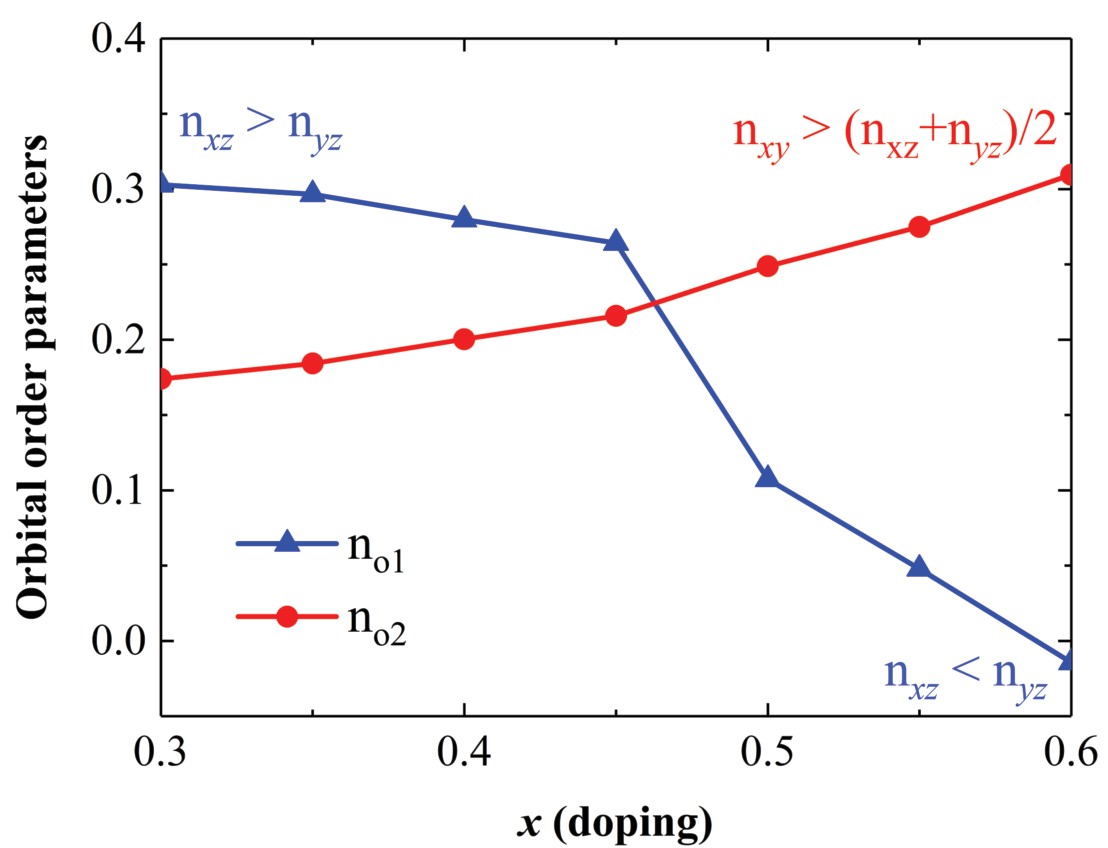}
\caption{(Color online) Dependence of OO parameters on doping concentration $x$ (only valid for the AFM-II/SC-II phases with $x$ ranging from 0.3 to 0.6) within the five-orbital Hubbard model of the second parent phase LaFeAsO$_{0.5}$H$_{0.5}$ with $U$=1.7 eV and $J_{H}$=0.1$U$.}
\label{Fig3}
\end{figure}

To explore the orbital-spin ordering, we investigate the five-orbital Hubbard models within the mean-filed approximation for $x$=0 and $x$=0.5 situations. The order parameters $n_{\alpha}$ and $m_{\alpha}$ are defined in Eq.~\ref{eq.3}.
Our results suggest a weak orbital polarization $n_{o1}=n_{xz}-n_{yz}$ with $n_{xz}>n_{yz}$ at $x$=0. In contrast, there is a strong orbital polarization $n_{o2}=n_{xy}-(n_{xz}+n_{yz})/2$ with $n_{xy}>n_{xz}$ or $n_{yz}$ at $x$=0.5, as displayed in Fig.~\ref{Fig2} and~\ref{Fig3}.
This strong $d_{xy}$ ferro-OO can lead to a very large N.N. spin exchange parameter $J_{2}$, which explains the strong anisotropy between $J_{1}$ and $J_{2}$ \cite{PRL113-027002}. Therefore, the OO scenario involved with three quasi-degenerate orbitals in LaFeAsO$_{1-x}$H$_{x}$ ($x$=0.5) is significantly different from the weak OO in LaFeAsO ($x$=0).

We determine the magnetic ground state at the mean-field level by considering the following magnetic vectors $\mathbf{Q}$ including (0, 0), ($\pi$, 0), ($\pi$, $\pi$), ($\pi$, $\pi$/3) and (2$\pi$/3, 0). The consequence is that the SAFM phase with $\mathbf{Q}$=($\pi$, 0) is the most stable, in consistent with our preceding LDA data and the experimental results \cite{nphys10-300}. The dependence of the orbital occupancies and magnetic moments on Coulomb interaction $U$ are shown in Fig.~\ref{Fig2} (a) for $x$=0 and Fig.~\ref{Fig2} (b) for $x$=0.5, respectively. It is obviously found that in LaFeAsO$_{1-x}$H$_{x}$ at $x$=0.5 the orbital polarization occurs in three orbitals when the system enters into the second magnetic parent phase (AFM-II). The original quasi-degenerate three orbitals with almost equal particle number undergo a significant reconstruction, resulting in a slight polarization $n_{o1}=n_{xz}-n_{yz}$ ($n_{xz}<n_{yz}$) but a very large one $n_{o2}=n_{xy}-(n_{xz}+n_{yz})/2$. The large $n_{o2}$ indicates a strong $d_{xy}$ ferro-OO compared with the weak $d_{xz}$/$d_{yz}$-OO, as seen in Fig.~\ref{Fig2}. The strong $d_{xy}$ ferro-OO thus leads to a very large N.N. spin exchange parameter $J_{2}$, which explains the strong anisotropy between $J_{1}$ and $J_{2}$ \cite{PRL113-027002}. Therefore, the OO scenario involved with three quasi-degenerate orbitals in LaFeAsO$_{1-x}$H$_{x}$ ($x$=0.5) is significantly different from the weak OO in LaFeAsO at $x$=0.

To understand the doping dependence of the electronic properties of LaFeAsO$_{1-x}$H$_{x}$, we also solve the self-consistent equations of the five-orbital Hubbard model for different doping cases within the mean-field approximation. The doping dependence of the OO parameters is shown in Fig.~\ref{Fig3}, where $n_{o1}=n_{xz}-n_{yz}$ and $n_{o2}=n_{xy}-(n_{xz}+n_{yz})/2$.
These two OO parameters show a competing behavior: the $d_{xz}$-orbital weight increases while the $d_{xy}$-orbital weight decreases when $x$ varies from 0.5 to 0.35, and the emergence of SC-II is at $x$$<$0.5 end rather than at $x$$>$0.5 end. Out of expectation, when $x$$>$0.5, the $d_{xy}$-orbital weight becomes even more larger, consistent with the experimental observation \cite{nphys10-300,PRB91-064509}. These results explain why the SC-II phase appears at the left end of the AFM-II phase in the phase diagram of LaFeAsO$_{1-x}$H$_{x}$.

We previously used the methods beyond mean-field theory, {\it e.g.} slave-boson \cite{JPCM25-125601} and dynamical mean-field theory (DMFT) \cite{PRB98-195137} methods, to treat with the electronic correlation in the iron-based materials. For comparison, the mean-field results will generally overestimate the magnetic moment, while the methods beyond mean-field theory can provide more accurate magnetic moment. It is demonstrated that although the mean-field method generally overestimates the magnetic moment (the strength of the spin order), it can reasonably present the trend of the spin/orbital-order parameters and doping evolution with the electronic correlation strength. 
Note that the previous DMFT results \cite{RRB94-224511} for LaFeAsO$_{1-x}$H$_{x}$ show that the total magnetic moment $m$ $=$ 0.66 $\mu_{B}$ at $x$ $=$ 0, and $m$ $\sim$ 0.6 $\mu_{B}$ at $x$ $=$ 0.5, are in good agreement with the experiments. Especially, it also can reproduce the overall doping evolution of the magnetic order. However, the DMFT calculation illustrates that it still cannot properly present the emergence of the second SC phase (SC-II) as the second magnetic parent phase (AFM-II) being suppressed in the intermediate doping $x$. Moreover, the DMFT calculation found that $d_{xy}$ orbital becomes important in heavily doped AFM-II/SC-II phases, which is also verified by our mean-field results.

\subsubsection{Magnetic instability and superconducting pairing state: RPA method}
\begin{figure}[htbp]
\hspace*{-2mm}
\centering
\includegraphics[trim = 0mm 0mm 0mm 0mm, clip=true, width=0.65 \columnwidth]{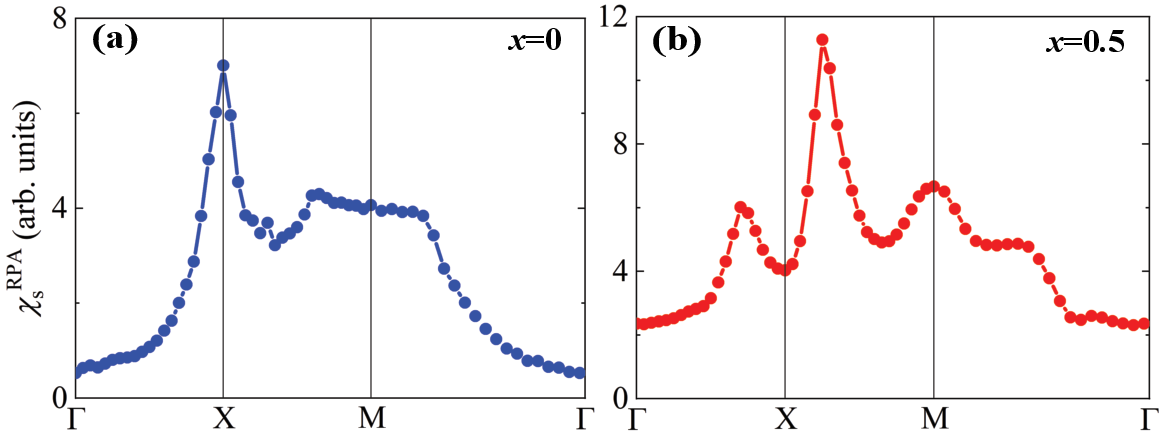}
\caption{(Color online) Dynamical spin susceptibilities of (a) LaFeAsO ($x$ $=$ 0) with $U$ $=$ 1.4 eV and $J_{H}$ $=$ 0.1$U$, and (b) LaFeAsO$_{1-x}$H$_{x}$ ($x$ $=$ 0.5)  with $U$ $=$ 0.9 eV and $J_{H}$ $=$ 0.1$U$.}
\label{Fig4}
\end{figure}

In order to explore the role of Fermi surface nesting and the possible magnetic instability in LaFeAsO$_{1-x}$H$_{x}$, we study the dynamical spin susceptibility within RPA. The obtained dynamical RPA susceptibilities are plotted in Fig.~\ref{Fig4} (a) and (b) for $x$ $=$ 0 and 0.5, respectively. It is found that the dynamical spin susceptibility peaks at $Q\sim$($\pi$, $\pi/3$) for $x$ $=$ 0.5 rather than ($\pi$, 0) for $x$ $=$ 0, indicating the nesting scenario fails to predict the magnetic structure in this electron-doped compound~\cite{PRB88-041106}. This is due to the absence of the hole pocket at $\Gamma$ point like K$_{1-x}$Fe$_{2-y}$Se$_{2}$, in contrast to that in LaFeAsO. 

\begin{figure}[htbp]
\hspace*{-2mm}
\centering
\includegraphics[trim = 0mm 0mm 0mm 0mm, clip=true, width=0.8 \columnwidth]{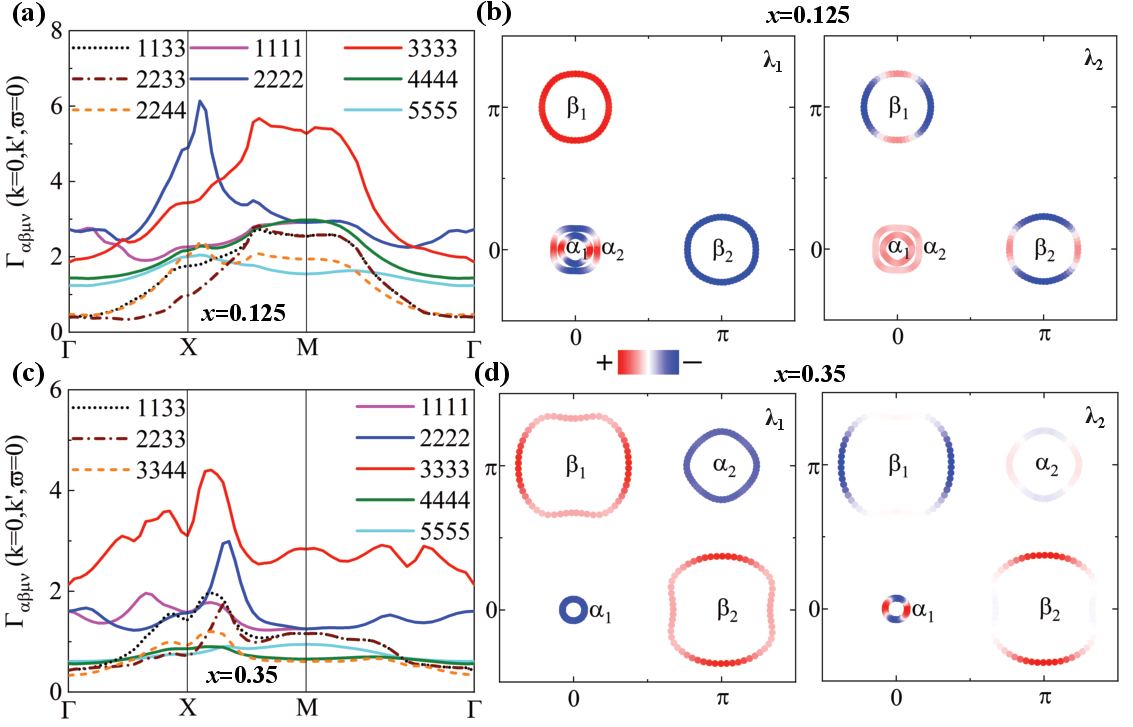}
\caption{(Color online) Orbital pairing vertices and two leading gap functions for LaFeAsO$_{1-x}$H$_{x}$, (a) and (b) $x$=0.125 with $U$=1.4 eV and $J_{H}$=0.1$U$, and (c) and (d) $x$=0.35 with $U$=0.8 eV and $J_{H}$=0.1$U$, respectively.}
\label{Fig5}
\end{figure}

Next, we present the effect of spin and orbital fluctuations on SC pairing in LaFeAsO$_{1-x}$H$_{x}$ within the RPA.
As shown in Fig.~\ref{Fig5}(a) and~\ref{Fig5}(c), the in-plane anisotropic $d_{yz}$ (or isotropic $d_{xy}$) orbital dominates the pairing vertex peaking around ($\pi$, 0) (or ($\pi$, $\pi$)) in SC-I phase, while the in-plane isotropic $d_{xy}$-orbital dominates over the entire Brillouin zone (BZ) in SC-II phase, implying an orbital-selective pairing state which is in line with the observation in iron chalcogenides by a recent ARPES experiment \cite{ncomms6-7777}. Notice that the off-diagonal elements, {\it i.e.} interorbital pairing vertices, mainly from orbital fluctuations, are considerably enhanced at $x$=0.35 compared with those at $x$=0.125. Therefore the orbital fluctuations due to the three quasi-degenerate orbitals also strongly contribute to the SC-II phase \cite{PRL112-187001}. The SC gap functions for both SC-I and SC-II phases are shown in Fig.~\ref{Fig5}(b) and~\ref{Fig5}(d), respectively. 
Although the topology of Fermi surface is so different for these two distinct SC phases \cite{PRL106-187001,PRB88-041106,PRB93-195148}, they remain relatively good nesting, suggesting a possibly enhanced spin and orbital fluctuations mediated pairing state.

The SC pairing symmetry is an important issue. In our RPA calculation as shown in Fig.~\ref{Fig5}, it is found that a $d_{x^{2}-y^{2}}$-wave state slightly dominates over or closely competes with an extended $s$-wave one for LaFeAsO$_{1-x}$H$_{x}$ at $x$ $=$ 0.125 (SC-I phase) with $U$=1.4 eV and the fixed $J_{H}$=0.1$U$, which indeed can be naturally explained as a consequence of the coexistence of ($\pi$, 0) and ($\pi$, $\pi$/2) spin fluctuations as asserted in Refs.~\cite{NJP11-025016,PRL101-087004}.
As clearly illustrated in our RPA calculation \cite{NJP11-025016,PRL101-087004}, the $s$-wave pairing state can dominate only for the very weak Hund’s rule coupling $J_{H}$ $=$ 0 at $x$ $=$ 0.125, though it remains nearly degenerate with $d$-wave pairing one.
In addition, for other lower doping $x$ $<$ 0.125, {\it e.g.} $x$ $=$ 0 and $x$ $=$ 0.1, the $s$-wave state is the most stable for most of the Hund’s rule coupling $J_{H}$. 
While for LaFeAsO$_{1-x}$H$_{x}$ at $x$ $=$ 0.35 (SC-II phase) with $U$=0.8 eV and the fixed $J_{H}$=0.1$U$, it is an $s_{\pm}$-wave state dominant over a $d_{x^{2}-y^{2}}$-wave one.  
Our results demonstrate that the SC pairing symmetry depends not only on doping level $x$, but also on interaction parameters $U$ and $J_{H}$.
For general reasons, we list at least two leading pairing symmetries. It is concluded that the $s$-wave symmetry is favored for most physical parameter cases in both SC-I and SC-II phases.
Now we turn to the spin/orbital fluctuations mediated SC pairing state. Within the RPA method based on the five-orbital Hubbard model, the pairing interaction includes the spin and orbital fluctuations channels in iron-based SC systems \cite{NJP11-025016,PRL112-187001}. Figure~\ref{Fig5} shows the inter-orbital and intra-orbital components of pairing vertices. The intra-orbital pairing vertices correspond to the contribution of spin fluctuations (solid line data), while the inter-orbital pairing vertices correspond to the contribution of orbital fluctuations (dotted line data), as seen in Fig.~\ref{Fig5}. It is found that the largest pairing vertices are the intra-orbital ones, which primarily affect the SC gap on the sections of the Fermi surface with the corresponding orbital character. Instead, most of the inter-orbital vertices are so small that can be neglected, only some inter-orbital components involved the active orbitals are considerable though not dominant, even relatively larger than the intra-orbital vertices related to the subdominant orbitals at certain $\mathbf{q}$ wave vectors. It can be demonstrated that the orbital-fluctuation mediated pairing is indeed weaker than that mediated by spin-fluctuation, but the active orbitals determine the pairing vertices character, implying an orbital selective scenario.
Actually, due to the relatively weak lattice distortion and moderate electronic correlation of most iron-based SC systems, the strength of spin order/fluctuation is generally greater than that of orbital order/fluctuation. Although orbital order/fluctuation is relatively weak, orbital degree of freedom (orbital-selective scenario) still has important effects on many aspects such as electronic states, magnetism, and SC pairing states.

It is worthy noting that the degeneracy or quasi-degeneracy of the on-site energies is not necessary to obtain strong spin and orbital fluctuations. More importantly, the weight of the degenerate or quasi-degenerate orbitals at the Fermi level and the good nesting are the essential elements for the strong spin and orbital fluctuations. In the orbital pairing vertices, the intra-orbital (inter-orbital) channels correspond to the spin (orbital) fluctuation, {\it i.e.} the peak intra-orbital (inter-orbital) vertices imply strong spin (orbital) fluctuation at some $\mathbf{q}$ vectors. As evident, in the systems covered in this work, the degenerate/quasi-degenerate related orbitals (active orbitals) do contribute significantly to the Fermi surface with relatively good nesting.
More importantly, all these degenerate/quasi-degenerate active orbitals contribute greatly to the pairing state as shown in the pairing vertices. Therefore, most degenerate orbitals already have maximum weight at Fermi level $E_{F}$, such as $d_{xz/yz}$, and $d_{xy}$ in LaFeAsO$_{1-x}$H$_{x}$ ($x$ $=$ 0 and 0.5) systems. On the other hand, the nesting characteristics of the Fermi surface are relatively evident, so there will be strong spin or orbital fluctuations.
The corresponding active orbitals contribute to the peak and sub-peak of the orbital pairing vertices, which proves the importance of the degenerate and quasi-degenerate orbitals.

Finally, the influence of orbital-fluctuation on pairing state is compared with the previous studies. In iron-based superconductor, within the RPA method, the spin-fluctuation mediated SC pairing state generally exhibits $s_{\pm}$-wave symmetry, while $s_{++}$-wave SC pairing state mediated by orbital-fluctuation can be obtained through including additional electron-phonon interaction \cite{PRB82-064518} or quadrupole interaction \cite{PRB85-134507,PRL112-187001} besides electron-electron interaction. Our RPA calculations obtain extended $s$ or $s_{\pm}$ wave pairing dominated by spin fluctuations, and the contribution of intra-orbital pairing vertices (corresponding spin-fluctuation contribution) are larger than that of inter-orbital pairing vertices (corresponding orbital-fluctuation contribution). Therefore, our calculated results do not show that the orbital fluctuation plays a leading role in pairing state, but only distinguish the dominant orbitals in the spin and orbital fluctuation-mediated pairing channels, indicating an orbital-selective mechanism.
In addition, our results are compared with the previous results of other more accurate methods to treat with the orbital fluctuations, such as DMFT \cite{RRB94-224511} and fluctuation exchange approximation (FLEX) \cite{PRL112-187001} calculations for LaFeAsO$_{1-x}$H$_{x}$. It is found that the orbital fluctuations indeed can be strong in LaFeAsO$_{1-x}$H$_{x}$, implying that our RPA calculation underestimate the contribution of orbital fluctuations, which deserves further investigation.

\subsubsection{Orbital-dependent phase diagram}
\begin{figure}[htbp]
\hspace*{-2mm}
\centering
\includegraphics[trim = 0mm 0mm 0mm 0mm, clip=true, width=0.65 \columnwidth]{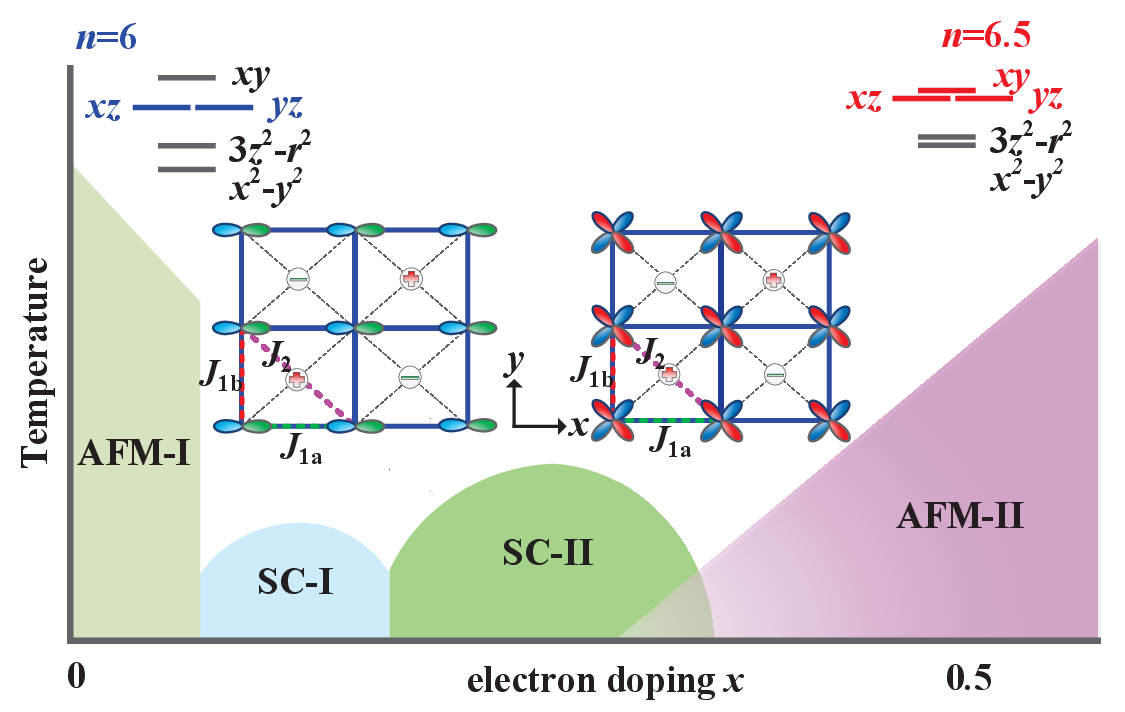}
\caption{(Color online) Schematic orbital-dependent phase diagram of LaFeAsO$_{1-x}$H$_{x}$ with the $d_{xz/yz}$ OO configuration at $x=0$, and the $d_{xy}$ OO configuration at $x=0.5$. Different orbital physics is displayed in AFM(I)-SC(I)-SC(II)-AFM(II) phase diagram with the first and second AFM/SC phases. The crystal field splittings for $x$=0 and 0.5 are displayed, respectively.}
\label{Fig6}
\end{figure}

Figure~\ref{Fig6} is a sketch of the orbital-dependent phase diagram of LaFeAsO$_{1-x}$H$_{x}$ with H doping based on our above analysis and a series of experiments reported \cite{ncomms3-943,nphys10-300,PRB91-064509}. As indicated above, we propose that two distinct AFM-SC phases, {\it i.e.} AFM(I)-SC(I) and AFM(II)-SC(II), are dominated by different orbital physics. For doping $x$ from 0 to 1, this phase diagram possesses AFM(I)-SC(I)-SC(II)-AFM(II) structure.
Note that the distribution form of the atomic orbital levels due to the crystalline field splitting of localized $d$ or $f$ orbitals surrounded by an anion ligand polyhedron corresponds to a topology of the crystalline field.
When a quasi-orbital degeneracy occurs in a multi-orbital system, it is usually unstable, and consequently a distinct parent phase (new structural phase or magnetic ordering one) appears, accompanied by a structural or magnetic phase transition.
Consequently, the origin of the second AFM/SC phases in LaFeAsO$_{1-x}$H$_{x}$ is induced by the quasi-degeneracy of the Fe-3$d$ orbitals.
This orbital scenario is different from that involved with only the $d_{xz}$ and $d_{yz}$ orbitals previously reported, indicating a completely novel orbital physics in the second parent (AFM-II) phase of LaFeAsO$_{1-x}$H$_{x}$.

\subsection{LaFeAs$_{1-x}$P$_{x}$O}
\subsubsection{Tight-binding model and electronic structures}
Up to date the crystal structures and atomic positions of LaFeAs$_{1-x}$P$_{x}$O at finite $x$ are not available. In order to investigate the doping dependence of the electronic properties of LaFeAs$_{1-x}$P$_{x}$O, we use the lattice parameters of the experimental crystal structure of LaFeAsO \cite{JACS130-3296} as the initial parameters to optimize the structural parameters of the $x$=0.5 phase (See Fig. S5 in the Supplementary Material \cite{SM}). Then, similar to the procedure of LaFeAsO$_{1-x}$H$_{x}$, the electronic structures, the projected Wannier functions (See Figs. S6 and S7 in the Supplementary Material \cite{SM}), and a five-orbital tight-binding model are all obtained based on the optimized structure for LaFeAs$_{0.5}$P$_{0.5}$O (LaFePO) at $x$=0.5 (1). The on-site energies of Fe-3$d$ orbitals for LaFeAs$_{1-x}$P$_{x}$O are also listed in Table.~\ref{Tab2}. For comparison, we list the on-site energies of three cases including LaFeAsO ($x$=0), LaFeAs$_{0.5}$P$_{0.5}$O ($x$=0.5) and LaFePO ($x$=1), respectively.
\begin{table}[htbp]
\caption{On-site energies of Fe-3$d$ orbitals for LaFeAs$_{1-x}$P$_{x}$O in unit of eV at $x$=0, 0.5 and 1, respectively.}
\label{Tab2}
\begin{tabular}{lccc}
\hline\hline
\multirow{1}{6cm}{LaFeAs$_{1-x}$P$_{x}$O}
$xz$/$yz$ & $xy$ & $x^{2}-y^{2}$ & $3z^{2}-r^{2}$ \\
\hline
\multirow{1}{6cm}{$x$=0 (LaFeAsO) \cite{PRL101-087004}}
$0.0$&$0.16$&$-0.34$&$-0.21$ \\
\hline
\multirow{1}{6cm}{$x$=0.5}
$0.0$&$0.31$&$\mathbf{-0.15}$&$\mathbf{-0.14}$\\
\hline
\multirow{1}{6cm}{$x$=1 (LaFePO) \cite{JPSJ79-044705}}
$0.0$&$0.35$&$-0.29$&$-0.20$ \\
\hline\hline
\end{tabular}
\end{table}
In LaFeAs$_{1-x}$P$_{x}$O, the isovalent substitution of P$^{3-}$ ions for As$^{3-}$ ions in the FeAs$_{1-x}$P$_{x}$ layer will lead to a compressed FeAs$_{1-x}$P$_{x}$ tetrahedra due to the smaller P$^{3-}$ ion radius than that of As$^{3-}$ ion, hence changing the crystal field splittings of Fe-3$d$ orbitals.
We performed a similar analysis for the two-dome case in LaFeAs$_{1-x}$P$_{x}$O, which is induced by the isovalent substitution. It is found that new quasi-degenerate $d_{3z^{2}-r^{2}}$ and $d_{x^{2}-y^{2}}$ orbitals in LaFeAs$_{1-x}$P$_{x}$O, different from the quasi-degenerate $d_{xy}$ and $d_{xz}$/$d_{yz}$ orbitals in LaFeAsO$_{1-x}$H$_{x}$, evidently occur at $x$$\sim$0.5, as shown in Table.~\ref{Tab2}.

\begin{figure}[htbp]\centering
\includegraphics[trim = 0mm 0mm 0mm 0mm, clip=true, width=0.65 \columnwidth]{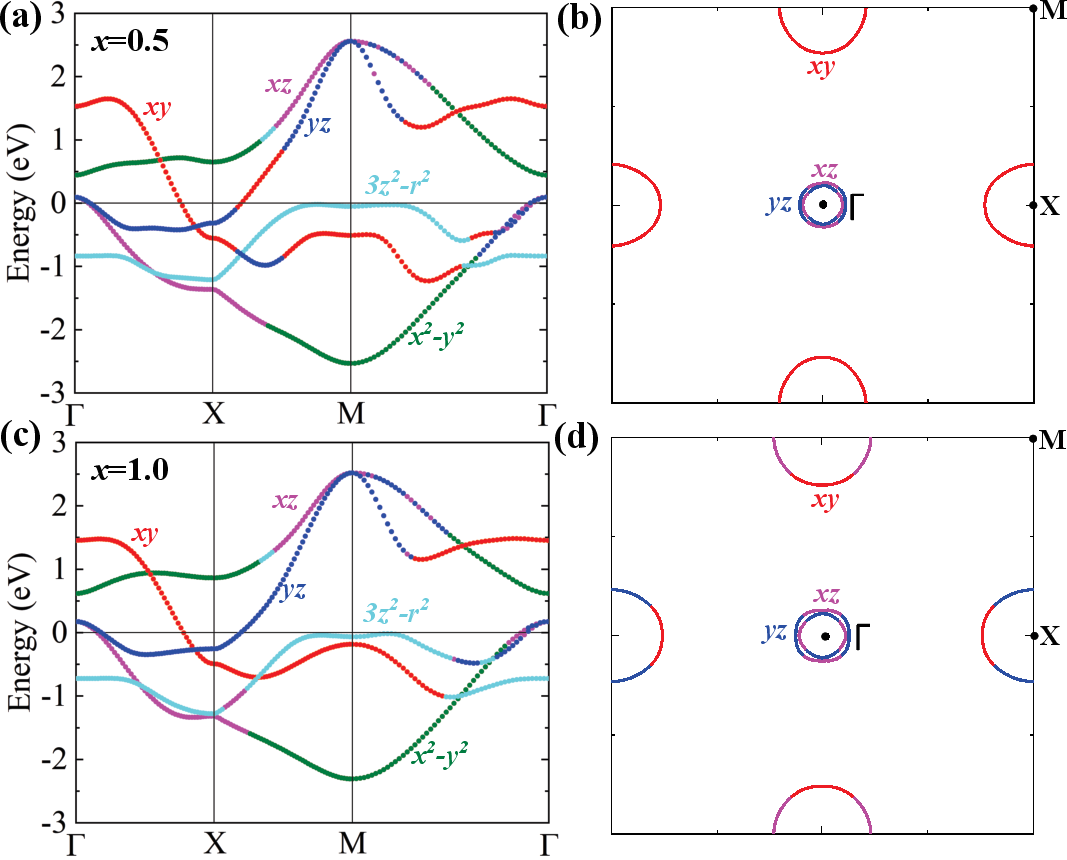}
\caption{(Color online) Band structures and Fermi surfaces of the five-orbital tight-binding model of LaFeAs$_{1-x}$P$_{x}$O at $x$=0.5 (a) and (b), and $x$=1.0 (LaFePO) (c) and (d), with colors indicating majority orbital character.}
\label{Fig7}
\end{figure}

Through constructing the five-orbital tight-binding model based on the projected Wannier functions of LaFeAs$_{0.5}$P$_{0.5}$O ($x$=0.5) and LaFePO ($x$=1), the obtained orbital-resolved band structures and Fermi surfaces are shown in Fig.~\ref{Fig7}. Note that our obtained electronic structures for $x$=1 are consistent with the experimental results in LaFePO \cite{PRL101-216402}.
Here we notice that in LaFeAs$_{1-x}$P$_{x}$O ($x$=0.5), one of the quasi-degenerate orbitals, $d_{3z^{2}-r^{2}}$ orbital does not appear at all on the Fermi surface with the dominant orbitals in Fig.~\ref{Fig7}(b). However, its orbital weight at Fermi level is relatively large and second only to the dominant orbitals, because it is not only extremely close to the Fermi level $E_{F}$, but also possesses flat-band characteristics.
In other words, the $d_{3z^{2}-r^{2}}$ orbital becomes an active one for the second parent phase LaFeAs$_{1-x}$P$_{x}$O ($x$=0.5).

\subsubsection{Magnetic instability and superconducting pairing state: RPA}
\begin{figure}[htbp]
\hspace*{-2mm}
\centering
\includegraphics[trim = 0mm 0mm 0mm 0mm, clip=true, width=0.65 \columnwidth]{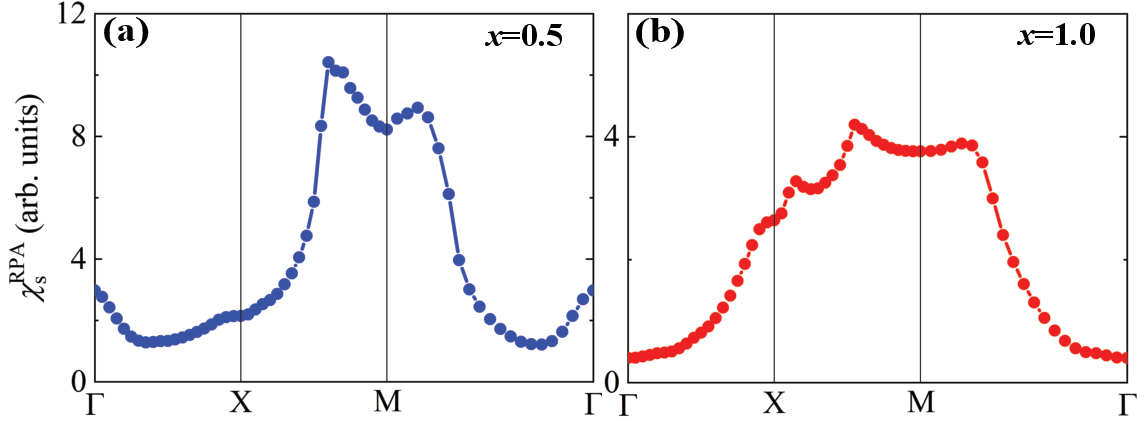}
\caption{(Color online) Dynamical spin susceptibilities of LaFeAs$_{1-x}$P$_{x}$O at $x$=0.5 (a) with parameters $U$=1.2 eV, and LaFeAs$_{1-x}$P$_{x}$O at $x$=1.0 (b) with parameters $U$=1.0 eV, respectively. The fixed parameter $J_{H}$=0.25$U$.}
\label{Fig8}
\end{figure}

The dynamical spin susceptibilities within RPA in LaFeAs$_{1-x}$P$_{x}$O are shown in Fig.~\ref{Fig8} (a) and (b) for $x$=0 and 0.5, respectively.
Obviously, the dynamical spin susceptibility peaks at $Q\sim$($\pi$, $\pi/2$) for $x$=0.5 (LaFeAs$_{0.5}$P$_{0.5}$O) and 1 (LaFePO) rather than ($\pi$, 0) for $x$=0, and displays a plateau around $Q\sim$($\pi$, $\pi$), implying different spin fluctuations in comparison with LaFeAsO ($x$=0).

\begin{figure}[htbp]\centering
\includegraphics[trim = 0mm 0mm 0mm 0mm, clip=true, width=0.8 \columnwidth]{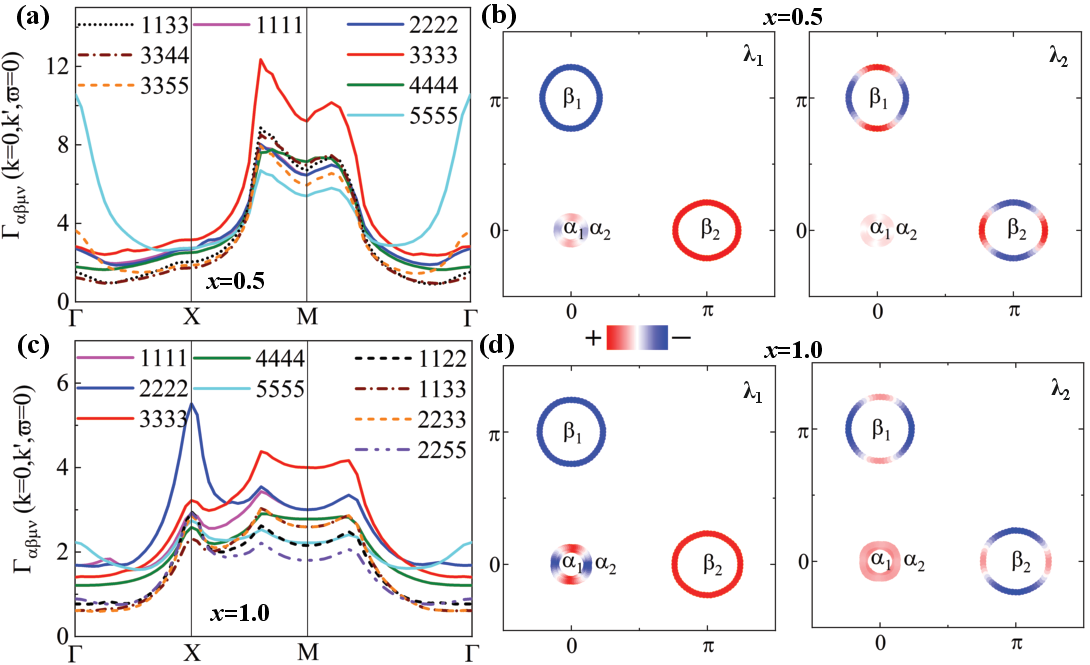}
\caption{(Color online) Orbital pairing vertices and two leading gap functions for LaFeAs$_{1-x}$P$_{x}$O at $x$=0.5 (a) and (b) with parameters $U$=1.2 eV and $J_{H}$=0.25$U$, and at $x$=1.0 (c) and (d) with parameters $U$=1.0 eV and $J_{H}$=0.25$U$.}
\label{Fig9}
\end{figure}

The SC pairings from the spin and orbital fluctuations in LaFeAs$_{1-x}$P$_{x}$O within RPA are displayed in Fig.~\ref{Fig9}.
Similarly, in addition to $d_{xy}$-orbital peaking at ($\pi$, $\pi$), $d_{3z^{2}-r^{2}}$ orbital favors (0, 0) spin fluctuations, and dominates the SC pairing state with the same pairing symmetry as SC-I, as shown in Fig.~\ref{Fig9}(a)-(d).
It is found that the active (near Fermi level $E_{F}$) in-plane isotropic $d_{3z^{2}-r^{2}}$-orbital dominates the AFM-II and SC-II phases, which is consistent with experimental reports \cite{JPSJ83-083702}. Especially, the P substitution of As contributes a flat band from the $d_{3z^{2}-r^{2}}$ orbital at Fermi level $E_{F}$, playing an essential role in the AFM-II and SC-II phases.

\subsubsection{Orbital-dependent phase diagram}
\begin{figure}[htbp]\centering
\includegraphics[trim = 0mm 0mm 0mm 0mm, clip=true, width=0.65 \columnwidth]{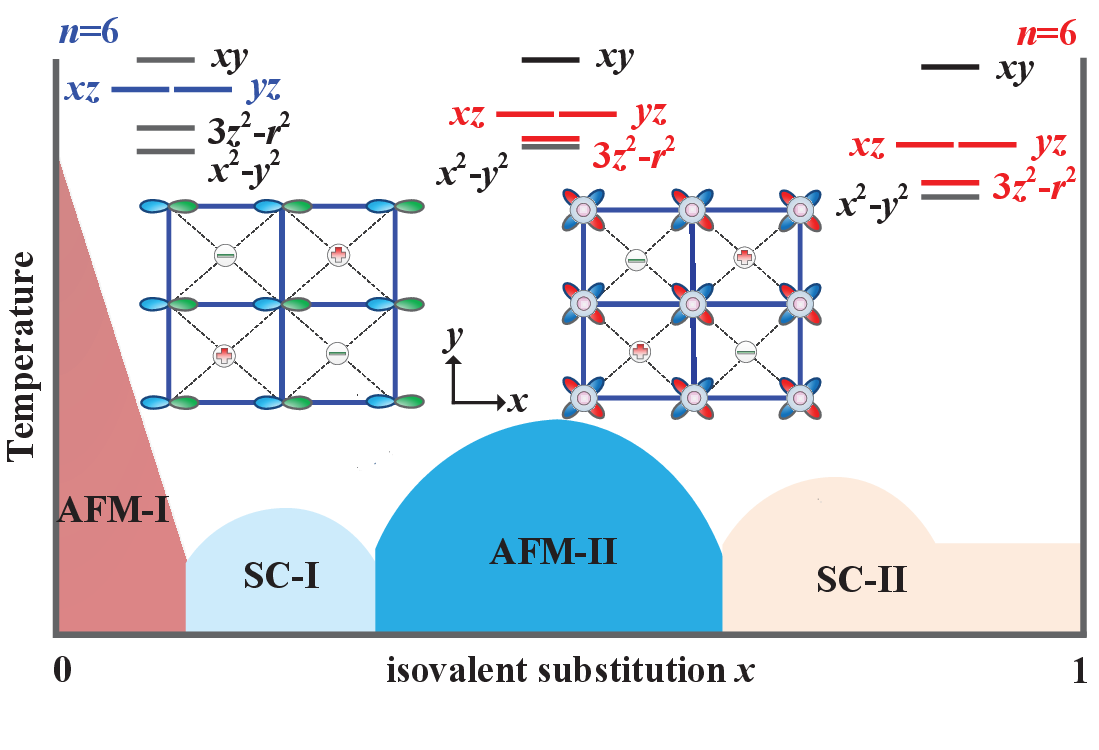}
\caption{(Color online) Schematic orbital-dependent phase diagram of LaFeAs$_{1-x}$P$_{x}$O with the $d_{xz/yz}$ OO configuration at $x=0$, and the $d_{xy}$ and $d_{3z^{2}-r^{2}}$ OO configuration at $x=0.5$. Different orbital physics are displayed in AFM(I)-SC(I)-AFM(II)-SC(II) phase diagram with the first and second AFM/SC phases. The crystal field splittings for $x$=0, 0.5 and 1 are displayed, respectively.}
\label{Fig10}
\end{figure}

Figure~\ref{Fig10} is a sketch of the orbital-dependent phase diagram of LaFeAs$_{1-x}$P$_{x}$O based on our theoretical analysis and experimental results \cite{JPSJ83-023707,JPSJ83-083702,PRB90-064504}. This phase diagram has a different dome structure with AFM(I)-SC(I)-AFM(II)-SC(II) from that of LaFeAsO$_{1-x}$H$_{x}$ with AFM(I)-SC(I)-SC(II)-AFM(II), demonstrating a completely different orbital physics.
Recently, one neutron diffraction experiment reported multiple magnetic orders in LaFeAs$_{1-x}$P$_{x}$O \cite{Communphys5-146}, demonstrating a competing magnetic and SC phase diagram.

\subsection{KFe$_{2}$As$_{2}$}
\subsubsection{Tight-binding model and electronic structures}
In exploring the electronic properties of KFe$_{2}$As$_{2}$ under pressure, we choose the experimental crystal structures and lattice parameters at $P$ = 0 \cite{PhysicaC469-332}, 10 \cite{arXiv1501.00330}, and 20 \cite{PRB91-060508} GPa, respectively. The obtained electronic structures at 0 GPa are consistent with our previous work \cite{RPB92-184512}. 
For the pressure of $P$ $=$ 30 GPa, due to the absence of the experimental data, we perform the structure optimization and obtain its optimized structure within the {\it first-principles} method. 
In fact, we have also verified the optimized structures of different pressures by comparing with the experimental results, concluding that there is a slight difference between them.  
In addition, the optimized lattice parameters of 30 GPa is confirmed to be consistent with the experimental extrapolated values \cite{PRB91-060508} and other calculated results \cite{JSNM33-2347}.
Then we obtain the electronic structures and the projected Wannier functions of both Fe-3$d$ and As-4$p$ orbitals for these different pressures (See Figs. S8-S11 in the Supplementary Material \cite{SM} for details). Also, the on-site energies of Fe-3$d$ and As-4$p$ orbitals for KFe$_{2}$As$_{2}$ under different pressures are obtained, and listed in Table.~\ref{Tab3}.
\begin{table}[htbp]
\caption{On-site energies of Fe-3$d$ and As-4$p$ orbitals for KFe$_{2}$As$_{2}$ under pressures of $P$ = 0 and 10 GPa in the tetragonal phase, and $P$ = 20 and 30 GPa in the collapsed tetragonal phase. The energy is measured in unit of eV.}
\label{Tab3}
\begin{tabular}{lccccc}
\hline\hline
\multirow{1}{4cm}{ }
$xz$/$yz$ & $xy$ & $x^{2}-y^{2}$ & $3z^{2}-r^{2}$ & $p_{x}$/$p_{y}$ & $p_{z}$ \\
\hline
\multirow{1}{4cm}{0 GPa}
$0.0$ &$-0.05$&$-0.51$&$-0.39$&$-1.33$&$-1.43$ \\
\hline
\multirow{1}{4cm}{10 GPa}
$\mathbf{0.0}$ &$\mathbf{-0.04}$&$-0.56$&$-0.41$&$-1.30$&$-1.41$ \\
\hline
\multirow{1}{4cm}{20 GPa (CT)}
$0.0$ &$\mathbf{-0.04}$&$-0.21$&$\mathbf{-0.05}$&$-1.79$&$-0.78$ \\
\hline
\multirow{1}{4cm}{30 GPa (CT)}
$0.0$ &$\mathbf{-0.03}$&$-0.21$&$\mathbf{-0.05}$&$-1.70$&$-0.83$ \\
\hline\hline
\end{tabular}
\end{table}
Notice that the hydrostatic pressure exerts an isotropic compression effect on the FeAs$_{4}$ tetrahedra similar to the chemical pressure mentioned above. It is found that in the tetragonal phase, the $d_{xy}$ orbital slowly becomes close to the $d_{xz}$/$d_{yz}$ orbitals under pressure, while in the collapsed tetragonal phase, the unfavored $d_{3z^{2}-r^{2}}$ orbital shifts up and eventually approaches the $d_{xy}$ orbital, indicating completely different quasi-degenerate orbitals in these two structural phases.
Another crucial feature in the CT phase, as opposed to the tetragonal phase, is that the $p_{z}$ orbital rises sharply toward the Fermi level $E_{F}$, implying a three-dimensional (3D) electronic state rather than a two-dimensional (2D) one in the tetragonal phase.

\begin{figure}[htbp]\centering
\includegraphics[trim = 0mm 0mm 0mm 0mm, clip=true, width=0.65 \columnwidth]{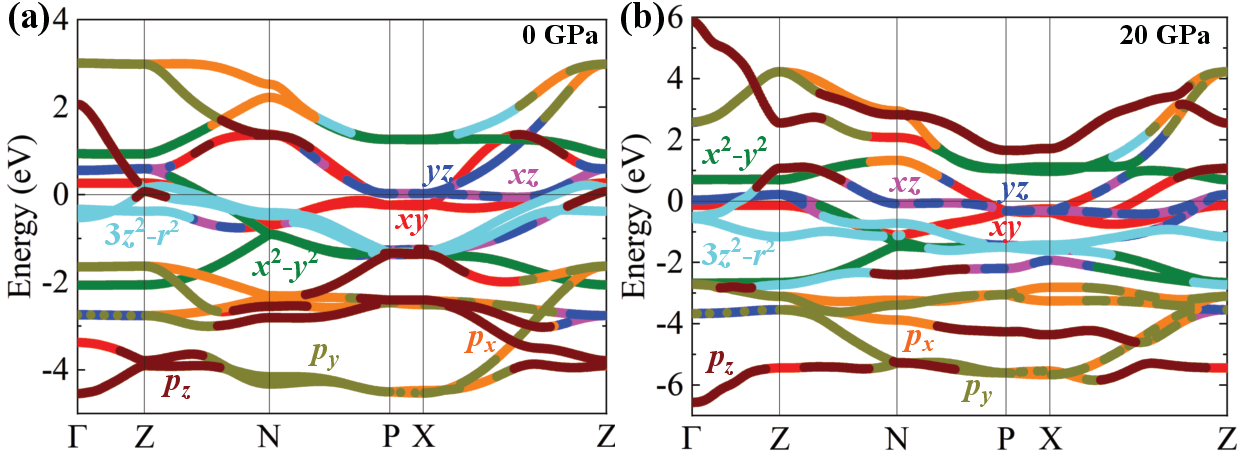}
\caption{(Color online) Band structures of KFe$_{2}$As$_{2}$ at (a) 0 GPa (tetragonal phase) and (b) 20 GPa (CT-phase) of the 3D $p$-$d$ models including 16 Wannier orbitals, with colors indicating majority orbital character.}
\label{Fig11}
\end{figure}

Accordingly, different from the 2D five-orbital model (10-band of two Fe atoms) of the tetragonal phase at 0 GPa, the 3D 16-band $p$-$d$ models are constructed due to large $p$-$d$ hybridization between Fe and As in the CT phase above 15 GPa. For comparison, the three-dimensional 16-band $p$-$d$ models of the tetragonal phase are also constructed. The orbital-resolved band structures of KFe$_{2}$As$_{2}$ at 0 GPa and 20 GPa are displayed in Fig.~\ref{Fig11}, which show two distinct electronic structures with sharply different orbital character.

\begin{figure}[htbp]\centering
\includegraphics[trim = 0mm 0mm 0mm 0mm, clip=true, width=0.65 \columnwidth]{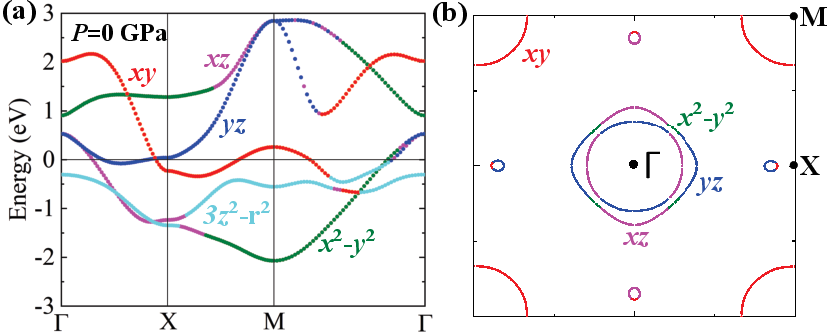}
\caption{(Color online) (a) Band structure and (b) Fermi surface of the five-orbital tight-binding model for KFe$_{2}$As$_{2}$ at $P$=0 GPa, with colors indicating majority orbital character.}
\label{Fig12}
\end{figure}

It is emphasized that due to the strong hybridization between the $p$ and $d$ orbitals under high pressure, the 2D five-orbital model fails to describe the CT phase (20 and 30 GPa). For simplicity, we only present the results of the low-pressure phase (0 GPa) focusing on the 2D 3$d$ orbital model in the following. The orbital-resolved band structure and Fermi surface of the five-orbital tight-binding model at $P$=0 GPa are plotted in Fig.~\ref{Fig12}.

\subsubsection{Superconducting pairing state: RPA}
\begin{figure}[htbp]\centering
\includegraphics[trim = 0mm 0mm 0mm 0mm, clip=true, width=0.65 \columnwidth]{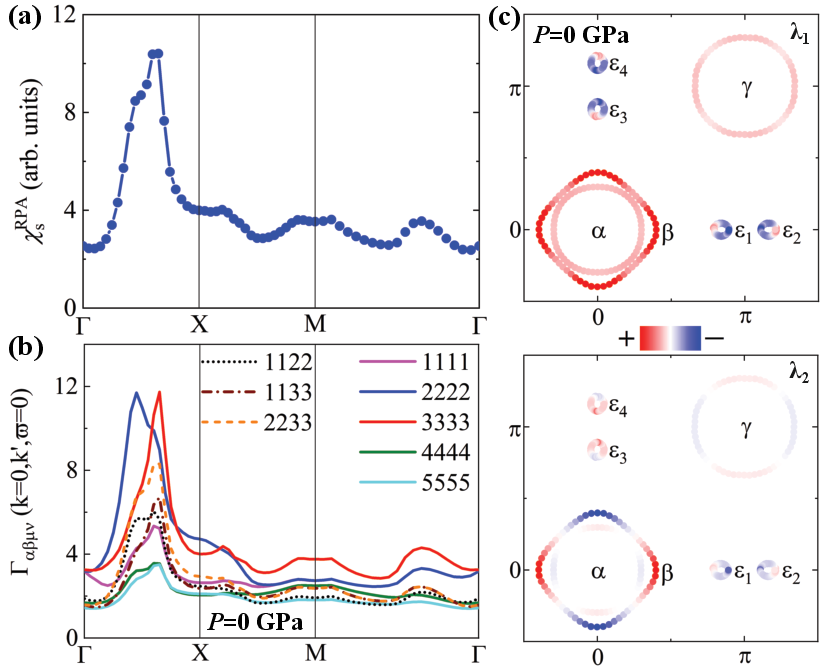}
\caption{(Color online) (a) Dynamical spin susceptibilities, (b) orbital pairing vertices, and (c) two leading gap functions of KFe$_{2}$As$_{2}$ at $P$=0 GPa with $U$=1.0 eV and $J_{H}$=0.25$U$.}
\label{Fig13}
\end{figure}

We have also analyzed the dynamical spin susceptibility and SC pairing vertice of KFe$_{2}$As$_{2}$ at 0 GPa, as shown in Fig.~\ref{Fig13}.
The dynamical spin susceptibility peaks at $Q\sim$($\pi/2$, 0)/(0, $\pi/2$) rather than ($\pi$, 0) for LaFeAsO. And orbital pairing vertices show a dominant intraorbital channel with the $d_{xz/yz}$ and $d_{xy}$ orbital characters.
That is, Figure~\ref{Fig13} shows that the SC-I pairing state with extended $s$-wave symmetry at 0 GPa is mainly dominated by $d_{xz/yz}$ and $d_{xy}$ orbitals at different wave vectors away from ($\pi$, 0) or ($\pi$, $\pi$), indicating a serious mismatch between spin and orbital fluctuations. This is possibly the reason for the low $T_{c}$ in tetragonal KFe$_{2}$As$_{2}$.

\subsubsection{Orbital-dependent phase diagram}
\begin{figure}[htbp]\centering
\includegraphics[trim = 0mm 0mm 0mm 0mm, clip=true, width=0.65 \columnwidth]{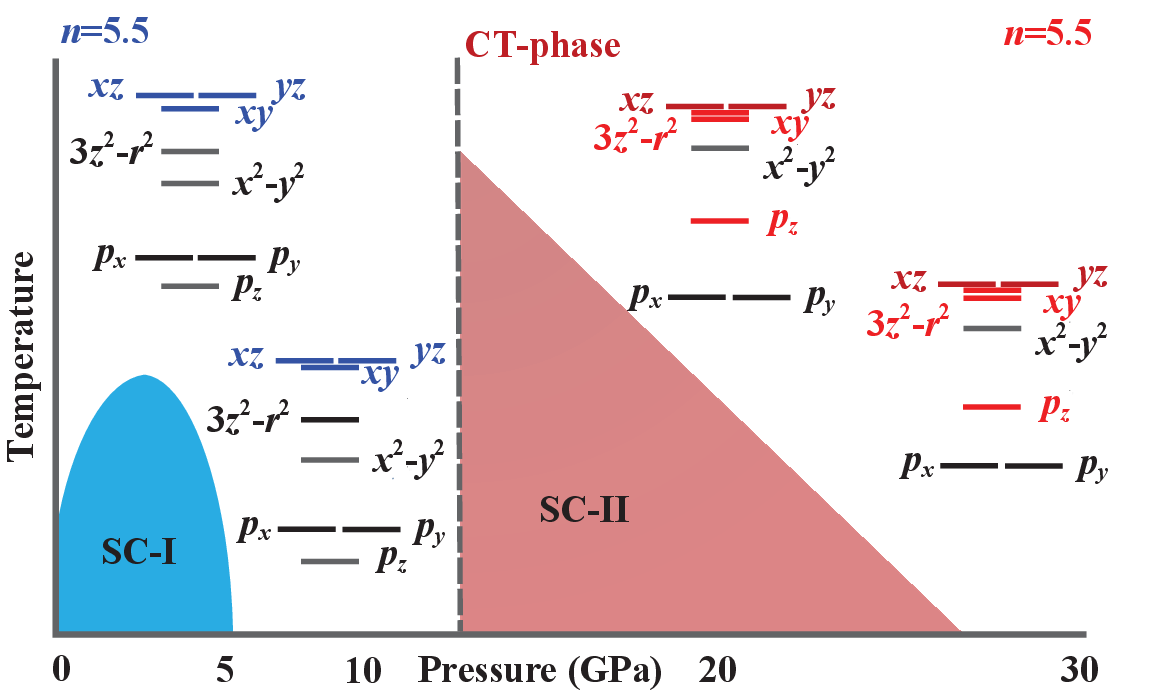}
\caption{(Color online)  Schematic orbital-dependent phase diagram of KFe$_{2}$As$_{2}$ under pressure. Different orbital physics are displayed in SC(I)-SC(II) phase diagram with the first and second structural/SC phases. The crystal field splittings for $P$=0, 10, 20 and 30 Gpa are displayed, respectively.}
\label{Fig14}
\end{figure}

Figure~\ref{Fig14} is a sketch of the orbital-dependent phase diagram of KFe$_{2}$As$_{2}$ under pressure based on our theoretical analysis and related experiments \cite{PRB91-060508}. This phase diagram shows an SC(I)-SC(II) dome structure separated by different structural phases, with no magnetic ordering phase due to heavy hole doping.
As shown in Fig.~\ref{Fig14}~\cite{PRB91-060508}, the quasi-degeneracy of orbitals takes place when the SC-I disappears or SC-II emerges.
The instability of the quasi-degenerate $d_{xy}$/$d_{xz/yz}$ orbitals at about 10 GPa drives a structural phase transition from tetragonal to CT under high pressure up to 15 GPa. The dominant $d_{xy}$-orbital character is presented at 10 GPa, but absent at 20 GPa, which is also verified by recent DFT and DMFT calculations \cite{PRB91-140503}.
It can be seen that the tetragonal and CT phases display completely distinct orbital physics. Although the active orbitals are different, the quasi-degenerate orbitals play an essential role in electronic and SC states in both the tetragonal and CT phases.

\subsection{Unified orbital-driven scenario and orbital-spin modes matching rule}
\begin{figure}[htbp]\centering
\includegraphics[trim = 0mm 0mm 0mm 0mm, clip=true, width=0.65 \columnwidth]{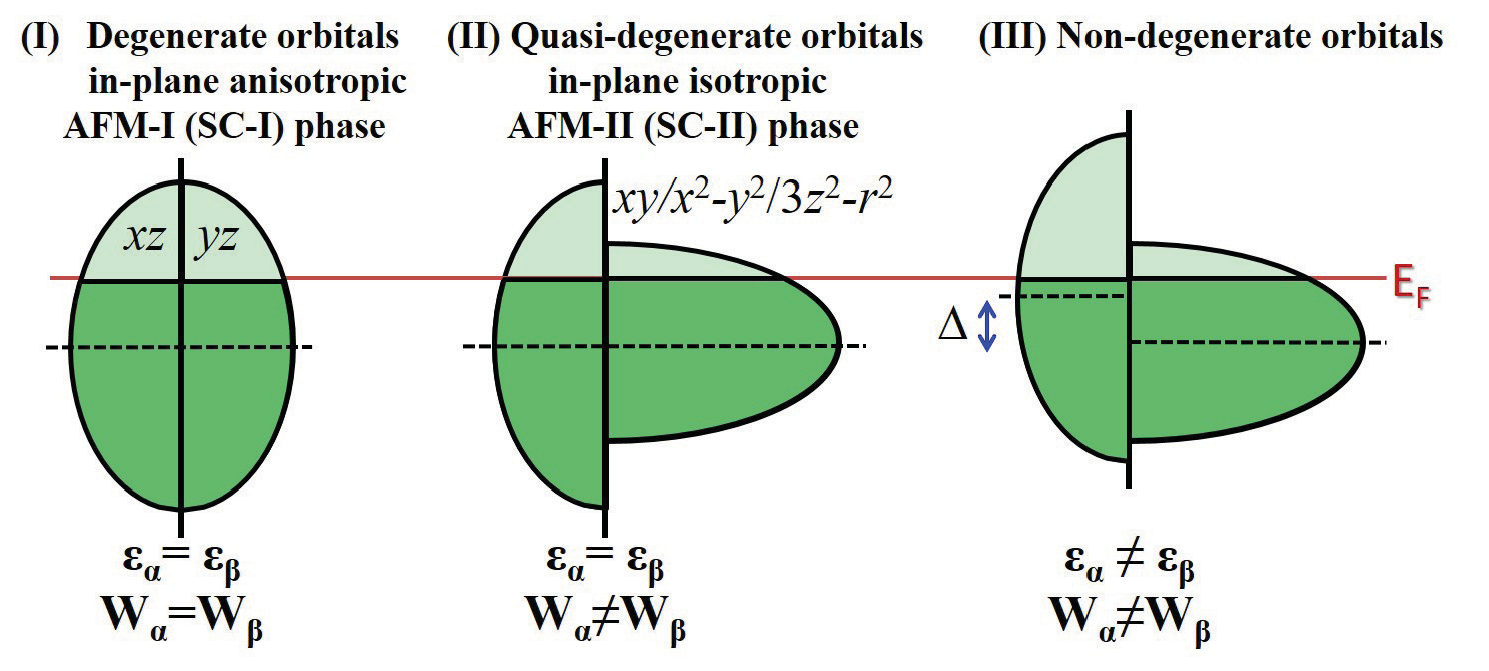}
\caption{(Color online) Sketch of three kinds of distinct orbital physics scenarios, {\it i.e.} full-degenerate orbitals, quasi-degenerate orbitals, and non-degenerate orbitals. $\alpha$ and $\beta$ represent two different ones of the five 3$d$ orbitals.}
\label{Fig15}
\end{figure}

In Fig.~\ref{Fig15}, we depict three kinds of distinct orbital physics scenarios. Conclusively, the active quasi-degenerate orbital drives the emergence of the AFM-II/SC-II phase, while the fully degenerate $d_{xz/yz}$-orbital dominates the AFM-I/SC-I phase.
Indeed, the doping/substitution or hydrostatic pressure pushes the in-plane anisotropic $d_{xz/yz}$-orbital with $C_{2}$ symmetry away from $E_{F}$, but it activates the in-plane isotropic $d_{xy}$ or $d_{x^{2}-y^{2}}$ with $C_{4}$ symmetry or fully-isotropic $d_{3z^{2}-r^{2}}$ orbital.
The in-plane isotropic orbital would suppress the ($\pi$, 0) spin fluctuation, but enhance the ($\pi$, $\pi$) or (0, 0) mode. Consequently, the orbital mode originating from the strong quasi-degeneracy in the AFM-II/SC-II phase also plays an essential role, which is supported by a recent NMR experiment \cite{PRB94-161104}.
When the multi-orbital system is continuously regulated by doping or pressure, the relative positions of the 3$d$ orbital levels (orbital level structure) of a transition metal ion will continuously change, so there always exists a degenerate state when two different orbital levels cross, that is the so-called quasi-orbital degeneracy ({\it i.e.} two orbitals with the identical energy, but different bandwidth). Indeed, it is the redistribution of the orbital levels that can be regarded as the change of the topological structure of crystalline field splitting. When the relative positions of multiple orbitals within orbital level structures do not change, it can be called that the topology of the crystal field remains unchanged, which is generally confined by the lattice symmetry and local ligand environment ({\it e.g.} four anions around an iron ion form an ideal or distorted tetrahedron, FeAs$_{4}$ tetrahedron here) of the system. Noted that the change in the topology of crystalline fields does not necessarily correspond to one structural phase transition.

\begin{figure}[htbp]\centering
\includegraphics[trim = 0mm 0mm 0mm 0mm, clip=true, width=0.65 \columnwidth]{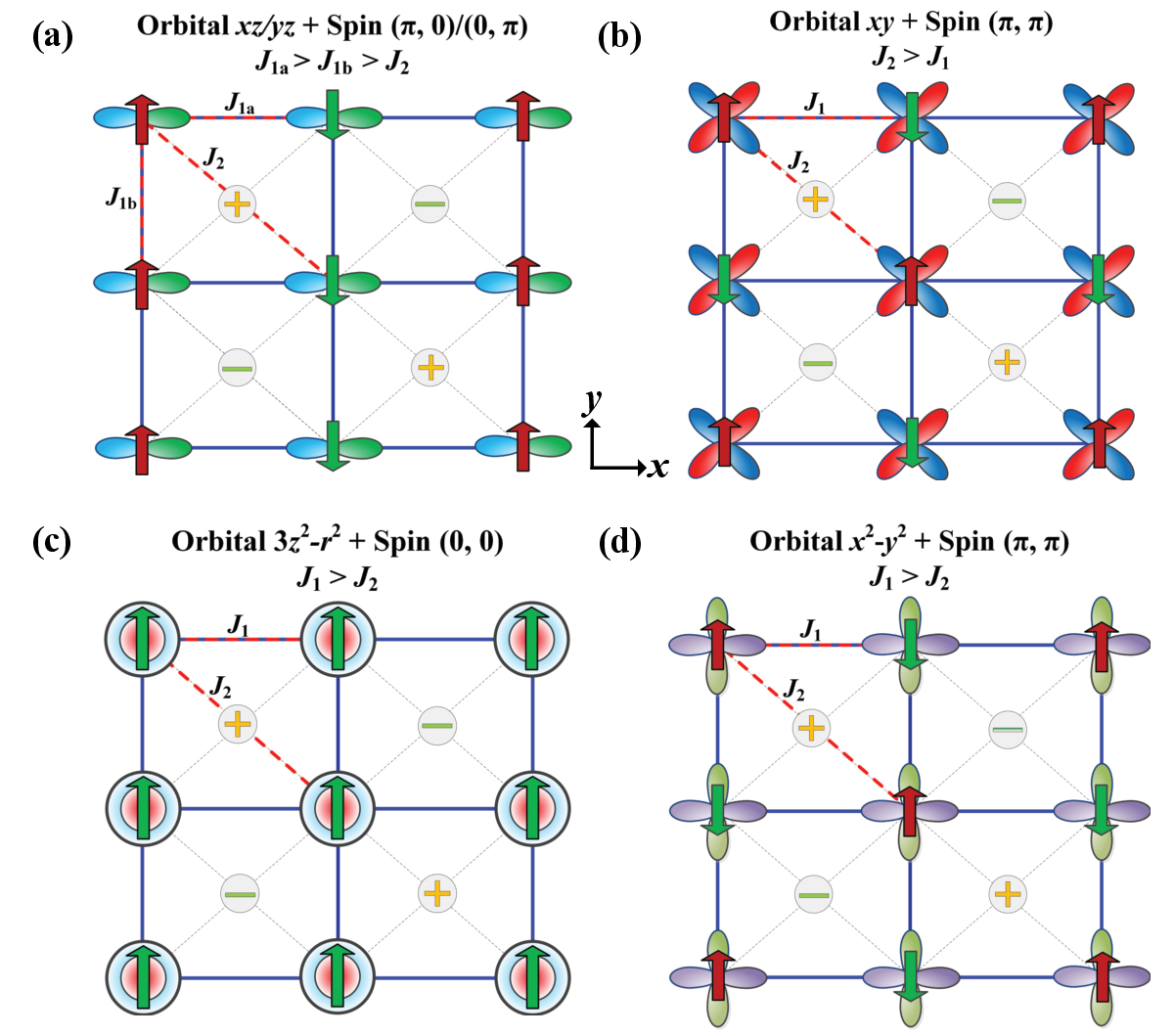}
\caption{(Color online) (a)-(d) Four types of ideal matching configurations of orbital-spin modes for transition-metal 3$d$ electrons.}
\label{Fig16}
\end{figure}

Moreover, similar to the Goodenough-Kanamori-Anderson (GKA) rule for the orbital-spin ordering in transition-metal oxides, we propose a matching rule of orbital-spin modes for high $T_{c}$ SC pairing symmetries in multi-orbital iron-based systems, as shown in Fig.~\ref{Fig16}(a)-(d), favoring high $T_{c}$ SC phases. Briefly, the $d_{xz/yz}$ ($d_{3z^{2}-r^{2}}$) orbital mode matches with ($\pi$, 0)/(0, $\pi$) ((0, 0)) spin mode with spin coupling $J_{1a}>J_{1b}$, and the $d_{xy}$ ($d_{x^{2}-y^{2}}$) orbital favors ($\pi$, $\pi$) spin mode with $J_{2}>J_{1}$ ($J_{1}>J_{2}$). Conversely, a mismatch may result in a relatively low-$T_{c}$ SC phase. Notice that the mixed pairing modes imply a competition or interplay among different orbitals, which is generally the case in real materials.
The dominant orbitals of the first and second SC (SC-I and S-II) phases have $C_{2}$ and $C_{4}$ symmetry respectively, and the higher symmetry of the latter may be the reason for its higher $T_{c}$.
In fact, one recent experiment reported that spin excitations are universally preferred by the orbital-selective SC pairing in iron-based superconductor CaKFe$_{4}$As$_{4}$ \cite{PRL128-137003}, supporting the orbital-spin fluctuations dependent pairing state.

To summarize, this two-dome phase is characterized by a large doping concentration variation $\Delta x$ $\sim$ 0.5, which is usually driven by orbital degeneracy or quasi-degeneracy, and accompanied by structural or magnetic phase transition.
By applying doping/substitution or high pressure, the energy increases for the system with the anisotropic $C_{2}$ symmetry corresponding to the fully degenerated $d_{xz}$/$d_{yz}$ orbitals, while decreases for the one with the isotropic $C_{4}$ symmetry corresponding to the quasi-degenerated orbitals, resulting in the emergence of the two distinctly different parent/SC phase characterized by the two-dome structure. As a consequence, an orbital-selective scenario generally exhibits in these two-dome AFM/SC phases of iron-based materials.

\section{Remarks and Conclusions}
Notice that there are two distinct types of multi-dome SC phases in iron-based SC systems as reported experimentally in FeAs-based materials and FeSe-based ones. Remarkably, the FeSe-based system is distinctly different from the FeAs-based one. Here we compare the different behavior of the doping-induced double-dome SC phase in the FeAs-based systems ({\it e.g.} LaFeAsO$_{1-x}$H$_{x}$, LaFeAs$_{1-x}$P$_{x}$O) with that in the FeSe-based systems ({\it e.g.} Li$_{x}$(Li,Fe)OHFeSe). In the FeAs-based system, there is usually a relatively strong electron-lattice coupling. Therefore, under doping (contributing carriers and chemical pressure), the spin and orbital degrees of freedom of electrons are closely coupled with the lattice, often resulting in new structural/magnetic phases. On the contrary, in the FeSe-based system, the electron-lattice coupling is weak, and the electron and lattice are generally decoupled under the intercalation doping, so the structural/magnetic phase transition is difficult to emerge. As a result, the two SC phases emerging in the system are purely originated from the electronic states, not structurally or magnetically related. Moreover, the Lifshitz transition of the electronic states can even induce the novel multiple SC phases according to our previous work \cite{NJP19-023028,PRB98-195137}. Especially, the FeSe system can undergo a significant change of electron-lattice coupling under substrate or high pressure, to produce new SC phases, as the experiments have reported.

Moreover, some SC phases, even if they have the same Fermi surface topology, can be different types of SC phases if the orbital characteristics are different. This solves the contradiction of Fermi surface nested scenario in explaining some systems with the same surface topology but completely distinct SC properties.
For comparison, there are completely different driving mechanisms in other systems besides the orbital and Lifshitz transition mechanisms in the iron-based system. For example, in twisted or Moir\'{e} graphene \cite{Nature556-43}, there are also multiple SC phases due to the flat-band mechanism.
We emphasize that the two-dome phase here is fine tuning with the particle occupancy change $\Delta n$ not exceeding 0.5, unlike these systems with different fillings, where $\Delta n$ $\sim$ 1.
It is emphasized that the two SC phases referred to here, are different from the those with $\Delta n$ $>$ 0.5, and even more different from that induced by flat bands from the twisted Moir\'{e} system. Here we mainly investigate the novel multiple SC cases when $n$ remains unchanged and $\Delta n$ $<$ 0.5.
The quasi-degenerate orbital scenario of the two-dome phases are completely different from those cases where the particle number varies greatly or the Moir\'{e} twist angle changes. Notice that the Moir\'{e} case also have multiple SC phases with unchanged particle number, indicating a similar multi-orbital/band character.

In summary, we have shown that in addition to conventional magnetic and SC phases in LaFeAsO$_{1-x}$H$_{x}$, LaFeAs$_{1-x}$P$_{x}$O and KFe$_{2}$As$_{2}$, the quasi-degenerate orbitals drive the emergence of the second AFM/SC phases, lead to universal two-dome SC phase in iron pnictides. Due to the orbital modulation, there exist two distinct types of SC phases, the in-plane anisotropic orbital dominates the SC-I phase with low $T_{c}$, the in-plane isotropic orbital is responsible for the SC-II phase with high $T_{c}$.
The physical essence of two-dome phase diagram is the redistribution of orbital crystal field (corresponding to the change of crystal field topology) under quantum manipulation of multi-orbital system, that is, multiple parent/SC phase dominated by different active orbitals.
From the symmetry perspective, quasi-degeneracy is due to the introduction of a new symmetry——quasi-symmetry or approximate symmetry (hidden symmetry), unlike full-degeneracy protected by crystal symmetry.
We note that this two-dome SC phase phenomenon has also been found in other iron-based SC materials, FeS under pressure \cite{npjQM2-49} and ThFeAsN$_{1-x}$O$_{x}$ \cite{JPCM30-255602}.
Our scenario could be generalized to similar two-dome SC phases observed experimentally in other iron-based SCs, such as the FeAs/FeSe/FeS-based compounds under pressure, as well as the heavy fermion SCs. The understanding of novel orbital-selective magnetic/SC state would shed light on the origin of the two-dome phases in iron-based materials. Our study suggests that, to search for iron-based SC materials with higher $T_{c}$, the presence of the active isotropic orbital and matching of orbital-spin fluctuations are the key factors.

\acknowledgements
This work was supported by the National Natural Science Foundation of China (NSFC) under Grant Nos. 11974354, 11774350, 11674296, 11574315, and 11474287. Numerical calculations were performed at the Center for Computational Science of CASHIPS.

\end{document}